\documentclass[twocolumn]{aastex63}
\usepackage{xcolor}

\shorttitle{High-Resolution Radio study of the Dragonfly Nebula}
\shortauthors{Jin et al.}
\graphicspath{{./}{figures/}}

\begin{document}

\title{High-Resolution Radio Study of the Dragonfly Pulsar Wind Nebula Powered by PSR J2021+3651}

\correspondingauthor{Ruolan Jin}
\email{ruolanjin@gmail.com}

\author[0000-0002-9652-0056]{Ruolan Jin}
\affiliation{Department of Physics, National Cheng Kung University, 70101 Tainan, Taiwan}

\author[0000-0002-5847-2612]{C.-Y. Ng}
\affiliation{Department of Physics, The University of Hong Kong, Pokfulam Road, Hong Kong}

\author[0000-0002-9396-9720]{Mallory S.E. Roberts}
\affiliation{Eureka Scientific}

\author[0000-0001-8229-2024]{Kwan-Lok Li}
\affiliation{Department of Physics, National Cheng Kung University, 70101 Tainan, Taiwan}

\begin{abstract}
The Dragonfly Nebula (G75.2$+$0.1) powered by the young pulsar J2021+3651 is a rare pulsar wind nebula (PWN) that shows double tori and polar jets enclosed by a bow-shock structure in X-rays. We present new radio observations of this source taken with the Very Large Array (VLA) at 6\,GHz. The radio PWN has an overall size about two times as large as the X-ray counterpart, consisting of a bright main body region in southwest, a narrow and fainter bridge region in northeast, and a dark gap in between. The nebula shows a radio spectrum much softer than that of a typical PWN. This could be resulting from compression by the ram pressure as the system travels mildly supersonically in the interstellar medium (ISM). Our polarization maps reveal a highly ordered and complex $B$-field structure. This can be explained by a toroidal field distorted by the pulsar motion.
\end{abstract}

\keywords{Unified Astronomy Thesaurus concepts: Pulsar wind nebulae (2215); Pulsars (1306); Radio interferometry (1346); Polarimetry (1278); Supernova remnants (1667)}

\section{Introduction} \label{sec:intro}
A pulsar has a strong electromagnetic field that can rip particles off the stellar surface and accelerate them to high energies. This drives a magnetized wind that are composed of mainly electrons and positrons. As the wind streams into the surrounding medium, the particles are shocked and then are further accelerated to ultra-relativistic energies. This results in a shocked wind bubble known as a pulsar wind nebula (PWN) \citep[see][for a review]{Gaensler2006}. These systems emit synchrotron radiation from radio to X-rays and inverse Compton (IC) scattering radiation from MeV to TeV gamma-rays. They serve as good nearby laboratories for studying relativistic shock physics and acceleration mechanism of Galactic cosmic rays. They are also suggested to be a main contributor to the detected positron excess in the Galaxy \citep{Blasi2011, DellaTorre2015}. 

Arcsecond resolution X-ray observations revealed small-scale structures in PWNe \citep{Kargaltsev2008, Kargaltsev2012} including axisymmetric tori and jets in subsonic moving PWNe \citep[see][]{Ng2004,NgRomani2008}. These are resulting from toroidal magnetic field and anisotropic (latitude-dependent) energy outflow as numerical simulations suggested \citep[e.g.,][]{Olmi2016, Porth2017}. A pulsar inside a supernova remnant (SNR) generally moves subsonically or transonically due to the high sound speed (hundreds of km\,s$^{-1}$) in the hot environment. Therefore, the PWN structure, such as the jets and tori, are not significantly perturbed. Once it leaves the SNR and enters the interstellar medium (ISM), the sound speed decreases ($\sim$ a few to a few tens of km\,s$^{-1}$) and the pulsar motion  (typically a few hundreds of km\,s$^{-1}$) becomes supersonic. The pulsar wind is then confined by the ISM ram pressure, resulting in a bow shock. A bow shock PWN has a compact head followed by a long tail that can extend up to a few parsecs \citep[see e.g.,][]{Kargaltsev2017, Reynolds2017}. 

One special case is the Dragonfly (G75.2+0.1). This system is powered by PSR J2021+3651, an energetic (spin-down luminosity $\dot E=3.4\times10^{36}$\,erg\,s$^{-1}$) and young (characteristic age $\tau_c=P/2\dot P=17.2$\,kyr) pulsar with a rotation period of 103.7\,ms. The pulsar emission has been detected in radio \citep{Roberts2002}, X-rays \citep{Hessels2004}, and $\gamma$-rays \citep{Halpern2008}. The source distance is controversial. While the pulsar dispersion measure implies over 10\,kpc \citep{Yao2017}, X-ray spectral study suggests a closer value of 3--4\,kpc \citep{VanEtten2008}. A later study obtained a even closer distance $D=1.8_{-1.4}^{+1.7}$\,kpc with 90$\%$ confidence interval, using both the X-ray spectrum and red-clump stars as standard candles and employing the extinction-distance relation along the pulsar line of sight \citep{Kirichenko2015}. We will adopt this value throughout our study.  This partly overlaps with the estimate by \citet{VanEtten2008} and is also consistent with $D\approx1$\,kpc based on the empirical $\gamma$-ray ``pseudo-distance" relation \citep{SazParkinson2010}.
 
On a large scale, the Dragonfly exhibits a cometary tail in X-rays and radio \citep[see][]{Aliu2014} that points to the general direction of the TeV source HAWC J2019+368 located $\sim$0.3\arcdeg\ away \citep{Albert2021}. The latter was suggested to be the pulsar birthsite \citep{Mizuno2017} because TeV emission is originated from IC scattering of less energetic particles. It therefore traces the long-term history of a PWN system. By modeling the spectrum measured with HAWC and \emph{Suzaku}, it was estimated that the true age of PSR J2021+3651 is $\sim$7\,kyr \citep{Albert2021}. 

Previous X-ray observations of the Dragonfly with the \emph{Chandra X-ray Observatory} found that the PWN has axisymmetric main body of $\sim20\arcsec$ long. This can be fit with a double tori model. There are also polar jets of $\sim30\arcsec$ long extending from the pulsar \citep{Hessels2004,VanEtten2008}. Altogether the overall morphology resembles a dragonfly. The only other system having such a double tori morphology is the Vela PWN \citep{Helfand2001}. While the pulsar has a relatively young characteristic age (17.2\,kyr), the host SNR has never been detected. X-ray observation shows a bow-shock like structure surrounding the PWN. This could indicate that the SNR has dissipated and the pulsar is traveling supersonically in the interstellar medium (ISM). This makes the Dragonfly a rare case that shows both bow-shock and torus-jet structure in X-rays \citep{Hessels2004, VanEtten2008}. 

The typical features such as wisps and tori have been found both in radio and X-rays for some PWNe \citep[e.g., the Crab Nebula;][]{Dubner2017}. Our study will supplement in understanding its origin by comparing these structures with simulated results \citep[e.g.,][]{Barkov2019, Olmi2019}. The radio PWN can also trace the flow of the particles over longer time scale because of the longer cooling time. In order to look for the radio counterparts of the X-ray structure, we analyzed new VLA 4--8\,GHz (C-band) data with arcsecond resolution comparable with \emph{Chandra} images along with archived 1--2\,GHz (L-band) in this study. We will compare the radio bow-shock structure with that observed in X-rays in more details than the previous study in \citet{Aliu2014}. The magnetic field is a critical factor in the particle acceleration process. The new data with full polarization measurements allow us to map the magnetic field structure of the nebula in details for the first time. With this new observation in radio, we will constrain the injected particle spectrum to test the different acceleration theoretical models. By combining radio and X-ray spectra we can estimate the B-field strength or age. The observations and data analysis are reported in Section~\ref{sec:observation} and results are presented in Section~\ref{sec:results}. Based on these results, we discuss the morphology, injected particle spectrum, polarization properties of the system in Section~\ref{sec:discussion}.

\section{Observations and Data Analysis}\label{sec:observation}

\subsection{C band} \label{subsec:vlaobservation}
We carried out new radio observations of the Dragonfly on 2017 November 2 and December 1 using the Karl G. Jansky Very Large Array (VLA) in the B array configuration. The observations were done in the C-band (4--8\,GHz) with full polarization for a total on-source time of 3.5\,hr. Data were taken in the continuum mode comprising 32 spectral windows with a total bandwidth of 4\,GHz. The shortest array spacing is 243\,m, corresponding to a largest angular scale of $\sim40\arcsec$ that our radio maps are sensitive to. 3C147 was chosen for calibrating the flux density scale as well as the bandpass by the observations. To determine the antenna gains, a secondary calibrator J2015+3710 was observed every 20 minutes \citep{Perley2017}.

The VLA data reduction was performed using the Common Astronomy Software Applications \citep[CASA;][]{2007ASPC..376..127M}. We first applied the VLA CASA Calibration Pipeline (Version 5.6.1-8) to flag the shadowed or not-on-source antennas. The edge channels of each spectral window and of each baseband are flagged. The pipeline then removed radio frequency interference (RFI). We solved for the multi-band cross-hand delays of the two basebands ($\sim$2--4\,GHz and $\sim$4--6\,GHz) separately. Additional RFI flagging was performed after the pipeline processing.

After flagging and calibration, we employed the task \texttt{tclean} to generate images at 6\,GHz in CASA. We adopted a weighting parameter of robust = 1.5 to optimize between the sidelobe suppression and reducing noise levels \citep{Briggs1995}. The threshold for deconvolution is 3 times of the rms flux density. After the first round of cleaning, we applied a clean mask on the image interactively to restrict the region to be cleaned. The final continuum 6\,GHz total intensity image has a circular restoring beam of FWHM 3\arcsec\ and rms noise of $\sim$4\,$\mu$Jy\,beam$^{-1}$, which is similar to the theoretical value.

To generate polarization maps, we first applied similar flagging and calibration procedures as above to the polarization calibrators, then determined the instrumental delay between the two polarization outputs. After that, we solved for the instrumental polarization (the frequency-dependent leakage terms, so-called ``D-terms") as well as the polarization position angle using 3C138 and 3C147 as calibrators respectively \citep{Perley2013}. A natural weighting (robust = 2.0) was used to minimize the noise level. We deconvolved the images in Stokes Q and U simultaneously and restored them with a Gaussian beam of FWHM=4\arcsec\ to boost the signal-to-noise ratio (S/N). The rms noise of the images is compatible with the theoretical levels of $\sim$3\,$\mu$Jy\,beam$^{-1}$. The polarized intensity and polarization angle maps were created using the Stokes Q and U images with the task \texttt{immath}. We employed the task \texttt{rmfit} to calculate the rotation measure (RM) map, the intrinsic polarization angle (PA0) map after correction for Faraday effect, and the corresponding error maps. 

Finally, we divided the whole 4\,GHz bandwidth into two equal subbands of 2\,GHz each, centering at 5 and 7\,GHz, to measure the spectral index. We applied the same imaging and deconvolution procedure as above and restore with a slightly larger circular beam of FWHM 4\arcsec to boost the S/N. 

\subsection{L band} \label{subsec:vlaLbandobservation}
We processed an archival L-band observation of the Dragonfly taken with the VLA. The data were obtained as part of a small survey of Galactic EGRET sources in the field around PSR J2021+3651 in the D array on 2000 July 20 and in the C array on 2002 November 2 at a central frequency of 1.47\,GHz with a bandwidth of 50\,MHz. The flux (primary) calibrator used was J2015$+$3710 and 3C286 ($=$J1331$+$3030) were chosen for phase and bandpass calibration. A three-point mosaic centered on the pulsar was used, with each pointing in the mosaic getting roughly 85\,minutes of exposure in C array and 22\,minutes in D array. Therefore the shortest spacing of the baseline was 35\,m, which correspond to a angular resolution of $\sim16\arcmin$ in the L-band. The ATNF MIRIAD package \citep{Sault1995} was used to reduce the data and generate the image. The \texttt{mossdi} clean algorithm was used for deconvolution. Two bright sources near the edge of the field of view were subtracted from the $u$-$v$ visibilities and multiple rounds of self-calibration were done. Mainly phase self-calibration was used, although a few iterations of amplitude calibration were also done. The final gain restored and gain corrected image has a $12.4\arcsec\times11.3\arcsec$ beam with a noise level of roughly 0.13\,mJy\,beam$^{-1}$ rms near the center of the image. 

\subsection{X-ray}\label{subsec:xraydata}
Previous X-ray studies of the Dragonfly show that its spectrum is sensitive to the choice of region \citep{VanEtten2008,Mizuno2017}. In order to perform multi-wavelength comparison at different regions, we reanalyzed archival \emph{Chandra} X-ray observations (ObsIDs 7603 and 8502) taken in 2006. The data were reprocessed using the script \texttt{chandra\_repro} script in Chandra Interactive Analysis of Observations (CIAO) \citep{Fruscione2006} version 4.13. Both datasets were merged by applying the script \texttt{merge\_obs} in CIAO after astrometric check. We extracted the source spectra from regions defined by the radio morphology and background spectrum from nearby source free regions. The X-ray spectra were created using the \texttt{specextract} script of CIAO. We grouped the energy bins with at least 20 counts each and fit the spectra from both datasets simultaneously with the package \texttt{Sherpa} \citep{Freeman2001}. We employed an absorbed power-law model with the column density fixed at 6.7$\times10^{21}$\,cm$^{-2}$ following previous studies. All the X-ray spectral analyses were made in the energy range of 0.5--8\,keV.

\section{Results}\label{sec:results}
\subsection{Radio intensity map} \label{subsec:radio_inten}
\begin{figure}[htbp]
\includegraphics[width=1.0\columnwidth]{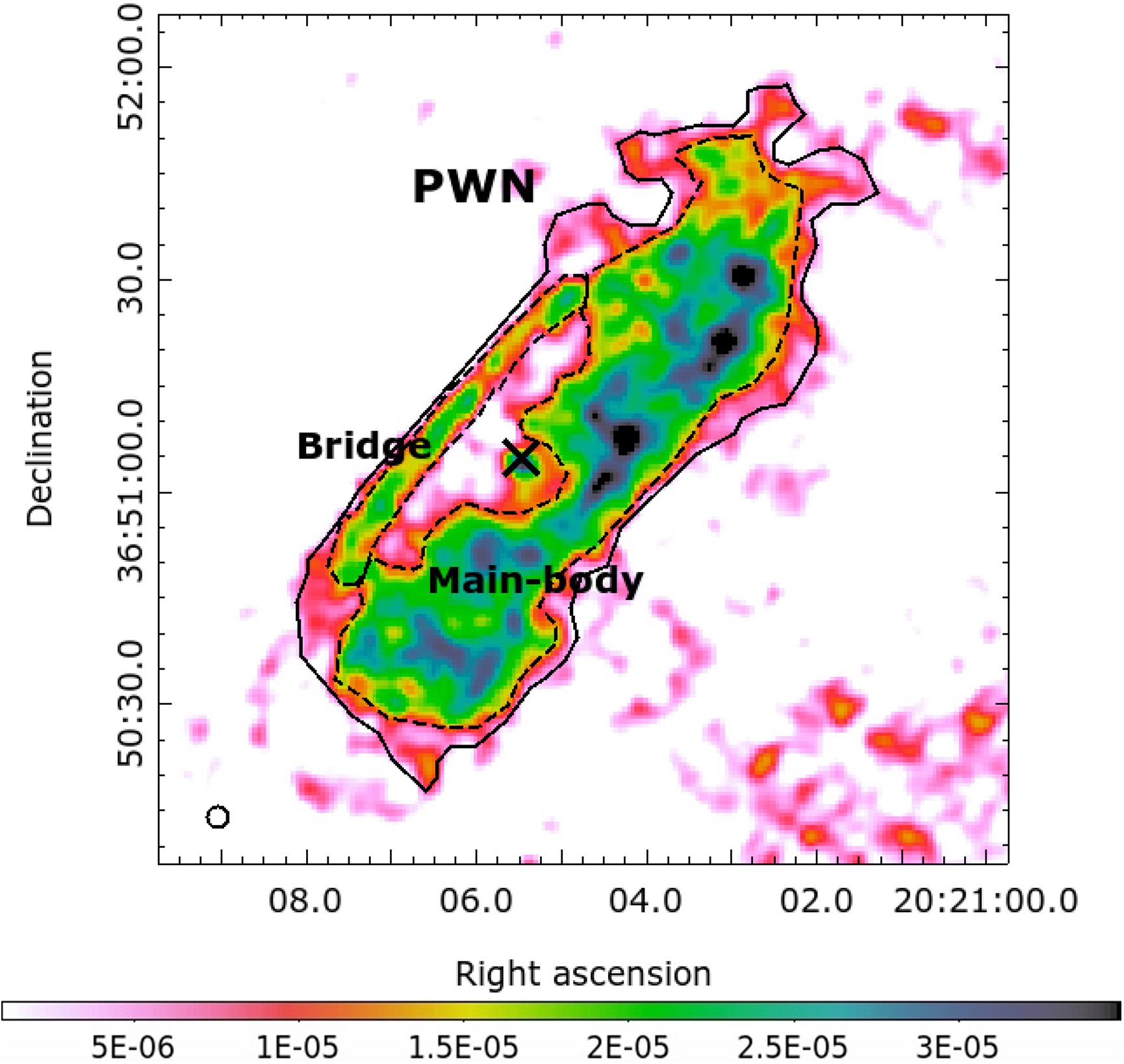}
\caption{Radio continuum image of the Dragonfly PWN at 6\,GHz clipped at 1\,$\mu$Jy\,beam$^{-1}$. The ``PWN" region is defined by the 10\,$\mu$Jy\,beam$^{-1}$ contour. The ``main-body" region and the ``bridge" are substructures of the PWN at the level of 15\,$\mu$Jy\,beam$^{-1}$. The position of PSR J2021+3651 is marked with an ``x". The beam size of FWHM $3\arcsec$ is shown on the lower left, and the color bar is in units of Jy\,beam$^{-1}$.}\label{fig:radio_intensity}
\end{figure}

\begin{figure}[htbp]
\gridline{\fig{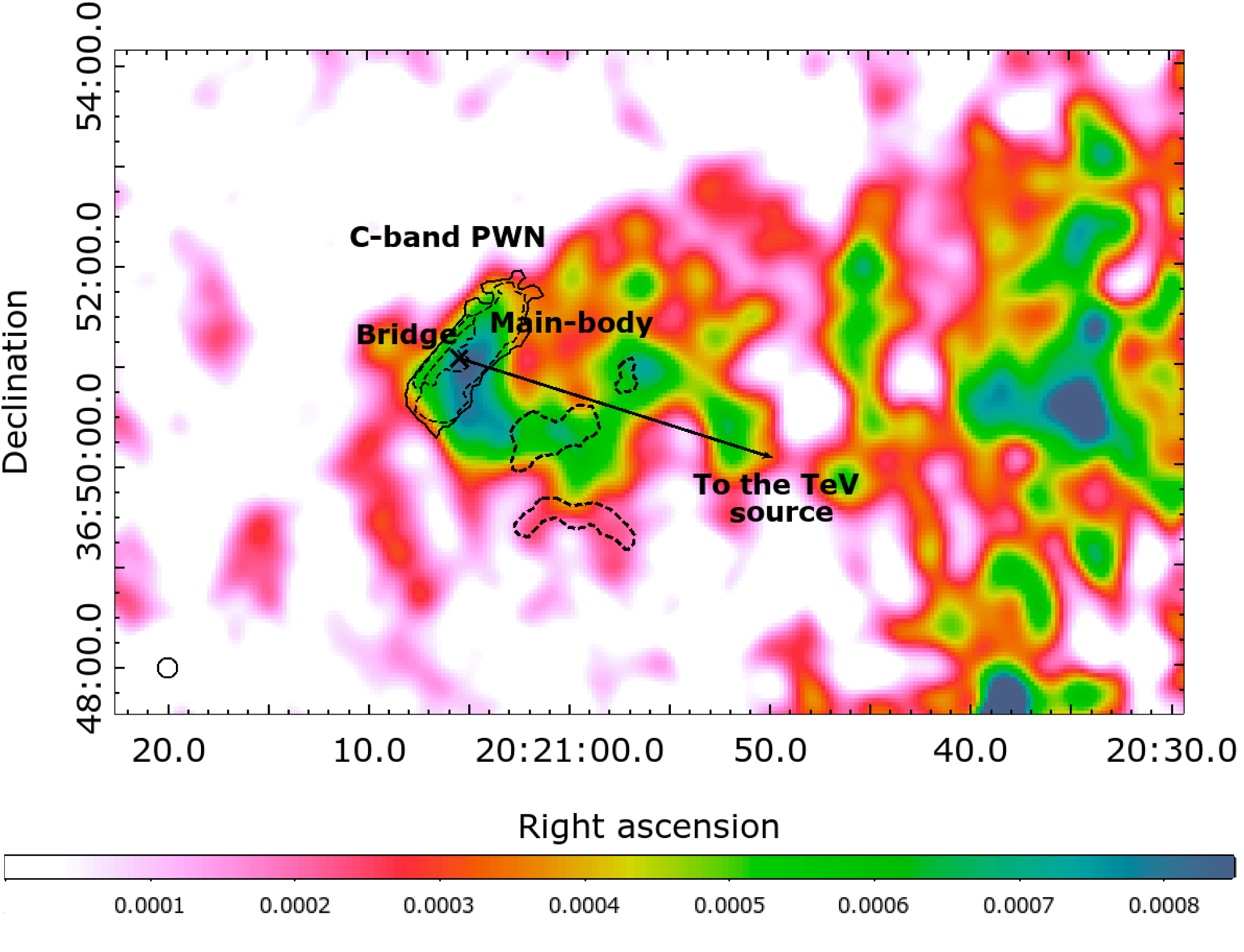}{1\columnwidth}{(a) L-band (1.5\,GHz) image}}
\gridline{\fig{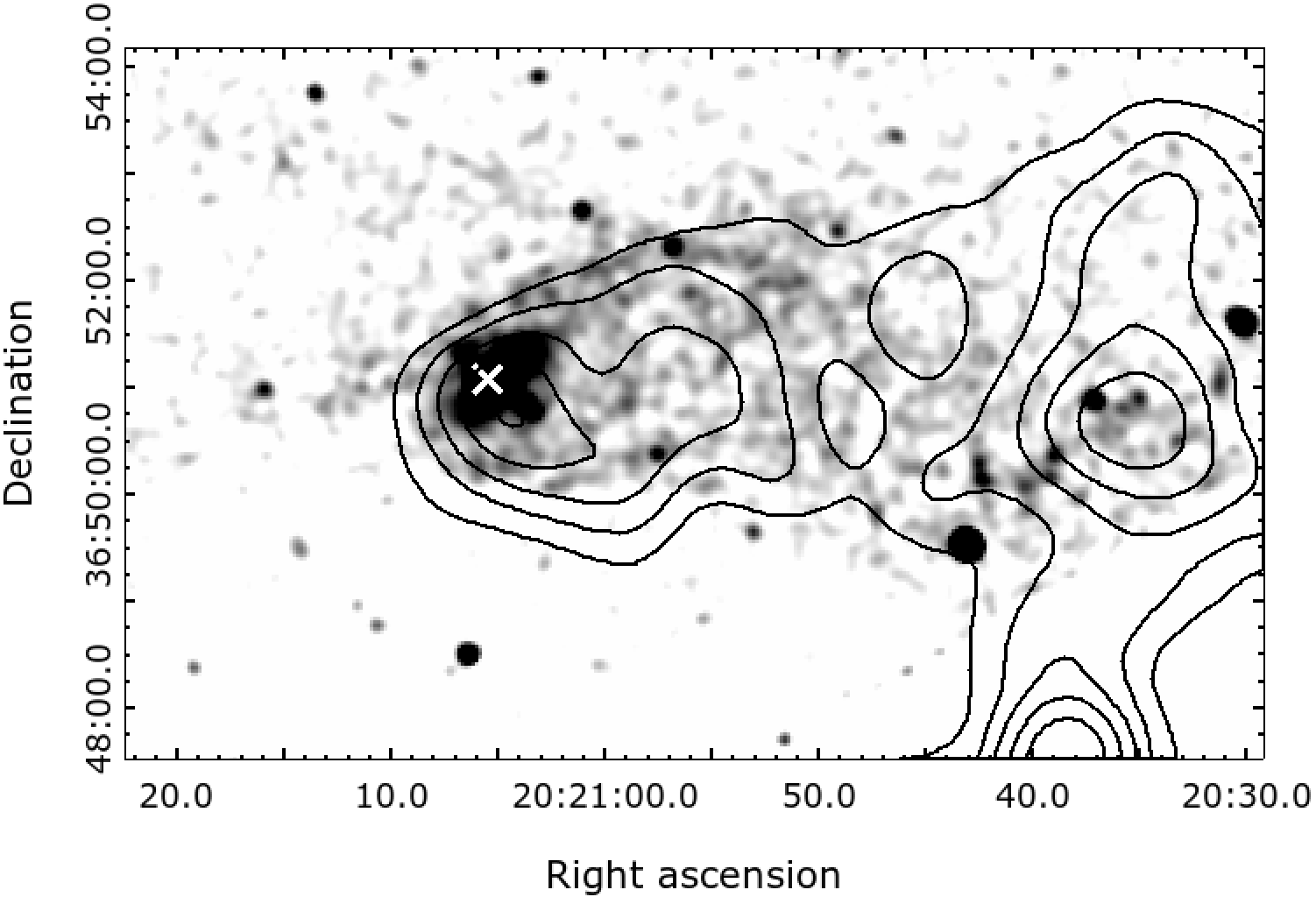}{1\columnwidth}{(b) X-ray image (0.3--7\,keV)}}
\caption{(a) Large scale radio emission surrounding the Dragonfly Nebula at the L-band (1.5\,GHz). The PWN regions and the emission lobes seen in the C-band are marked by solid and dashed contours respectively. The pulsar position is marked by x and the vector shows the direction from the pulsar to the pulsar's associated TeV source \citep{Albert2021}. (b) Smoothed \emph{Chandra} X-ray image (0.5--7\,keV) of the Dragonfly PWN, overlaid with the L-band intensity contours at 5, 10, 15, and 20\,$\mu$Jy\,beam$^{-1}$ levels.}\label{fig:zoomout}
\end{figure}

The radio intensity map of the Dragonfly at 6\,GHz is shown in Figure~\ref{fig:radio_intensity}. The PWN is clearly detected. It has a size of $1\farcm8\times0\farcm6$ and is elongated along a position angle (PA) of $\sim141\arcdeg$ (north to east). The overall structure consists of four parts: central resolved pulsar region, a western ``main-body" region, a fainter eastern ``bridge", and a dark gap lying between them. The surface brightness at the pulsar location is $\sim26$\,$\mu$Jy\,beam$^{-1}$. Since the FWHM of the central peak at the position of the pulsar is $\sim10\arcsec$ , which is larger than its beam size resolution (FWHM=3\arcsec), it should be a compact nebula surrounding the pulsar. The central pulsar region is surrounded by a faint ring-like region with $\sim$5\arcsec\ thickness. The latter has only half of the average surface brightness of the PWN. There are two brighter lobes southwest of the main PWN and another fainter one further south, which are marked by the dashed contours in Figure~\ref{fig:zoomout}a. 

The western main-body region dominates most of the radio flux density of PWN at 6\,GHz. It is $\sim$100\arcsec\ long and $\sim$20\arcsec\ wide and has a mean surface brightness of $\sim$20\,$\mu$Jy\,beam$^{-1}$. It contains two wings that are not perfectly symmetric about the central pulsar. The northwestern (NW) wing has length $\sim$60\arcsec, longer than the southeastern (SE) wing of $\sim$40\arcsec. The radio surface brightness is not uniform in the main-body region. The NW wing has higher peak brightness ($\sim$36\,$\mu$Jy\,beam$^{-1}$) than the SE one ($\sim$28\,$\mu$Jy\,beam$^{-1}$). Besides, $\sim$14\% of the area in the NW half has  surface brightness $\ge$ 25\,$\mu$Jy\,beam$^{-1}$, but only $<5\%$ in the SE wing, despite the mean C-band surface brightness in both wings are about the same (20\,$\mu$Jy\,beam$^{-1}$). 

On the eastern side of the PWN there is a relatively faint, long, and narrow bridge region with size $\sim$50\arcsec$\times$6\arcsec. The average surface brightness is $\sim$16\,$\mu$Jy\,beam$^{-1}$. In between the main-body and the bridge regions, we see a clear 5\arcsec\ wide gap. Its minimum intensity drops to $\sim$5\,$\mu$Jy\,beam$^{-1}$, which is only one-fifth of the peak emission at the pulsar position and is very close to the rms noise of 4\,$\mu$Jy\,beam$^{-1}$.  

The VLA L-band intensity map is shown in Figure~\ref{fig:zoomout}a with the C-band regions labelled. Overall, the cone-shape L-band emission extends from east to west over 10\arcmin. It shows two peaks, one at the center of the main-body and one at 6\arcmin\ west of the pulsar. The brightness is not uniformly distributed but rather patchy. It is unlikely due to the lack of short spacing because the largest angular size  that the observation is sensitive to is around 16\arcmin. The elliptical C-band emission is located at the head of the cone with $\sim$50\arcsec\ offset from the sharp edge in the east. The two brighter lobes southwest of the main PWN in the C-band image appear to correlate with the diffused L-band emission and another fainter patch is located near the edge of the L-band emission. 

\subsection{Multi-wavelength comparison} \label{subsec:multi_inten}
\begin{figure}[htbp]
\includegraphics[width=1.0\columnwidth]{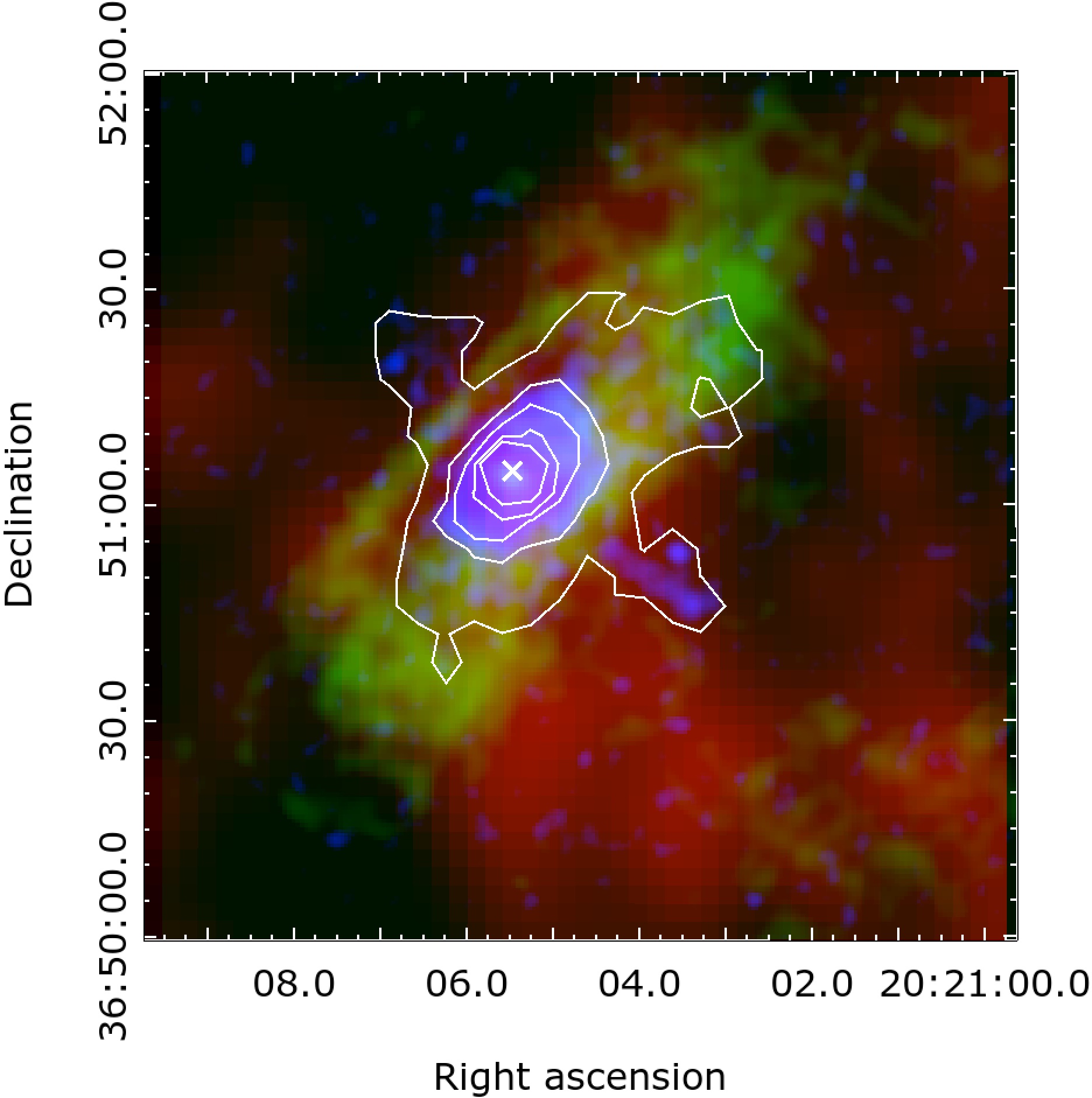}
\caption{3-color image of the Dragonfly Nebula. The L-band (1.5\,GHz), C-band (6\,GHz), and \emph{Chandra} (0.3--7\,keV) X-ray images are in red, green, and blue colors, respectively. The contours are from the 0.3--7\,keV X-ray image.}\label{fig:zoomin_3color}
\end{figure}

The cone-shape large scale radio structure shown in the L-band is also found in the X-rays (see Figure~\ref{fig:zoomout}b). The X-ray emission peaks at the pulsar while the peak in the L-band is offset to the south-west for $\sim$20\arcsec. Moving to a smaller scale, the elliptical radio C-band emission is about two times as large compared with the X-ray PWN as shown in Figure~\ref{fig:zoomin_3color}. The radio PWN's elongation aligns with the X-ray torus and is perpendicular to jets. The wings of the X-ray PWN lie within the radio C-band main-body while the X-ray jet partially overlaps with the radio bridge region. In X-rays, the NW wing, like its radio counterpart, is also slightly larger than its opposite half. We also found that the faint jet seen ahead of the bow shock in the Chandra image \citep{VanEtten2008} is pretty clear in the L-band (see Figure~\ref{fig:zoomout}). 

\begin{figure}[htbp]
\gridline{\fig{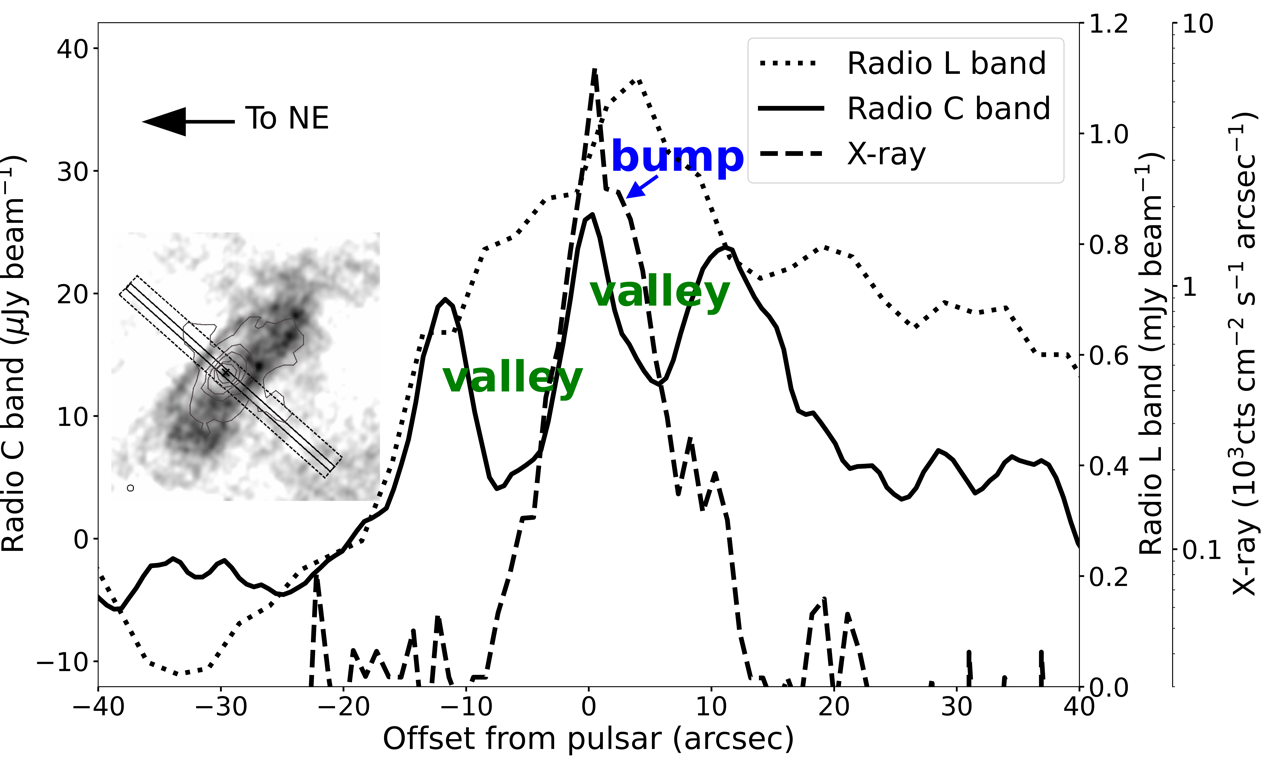}{1\columnwidth}{(a) Cross section along the minor axis}}
\gridline{\fig{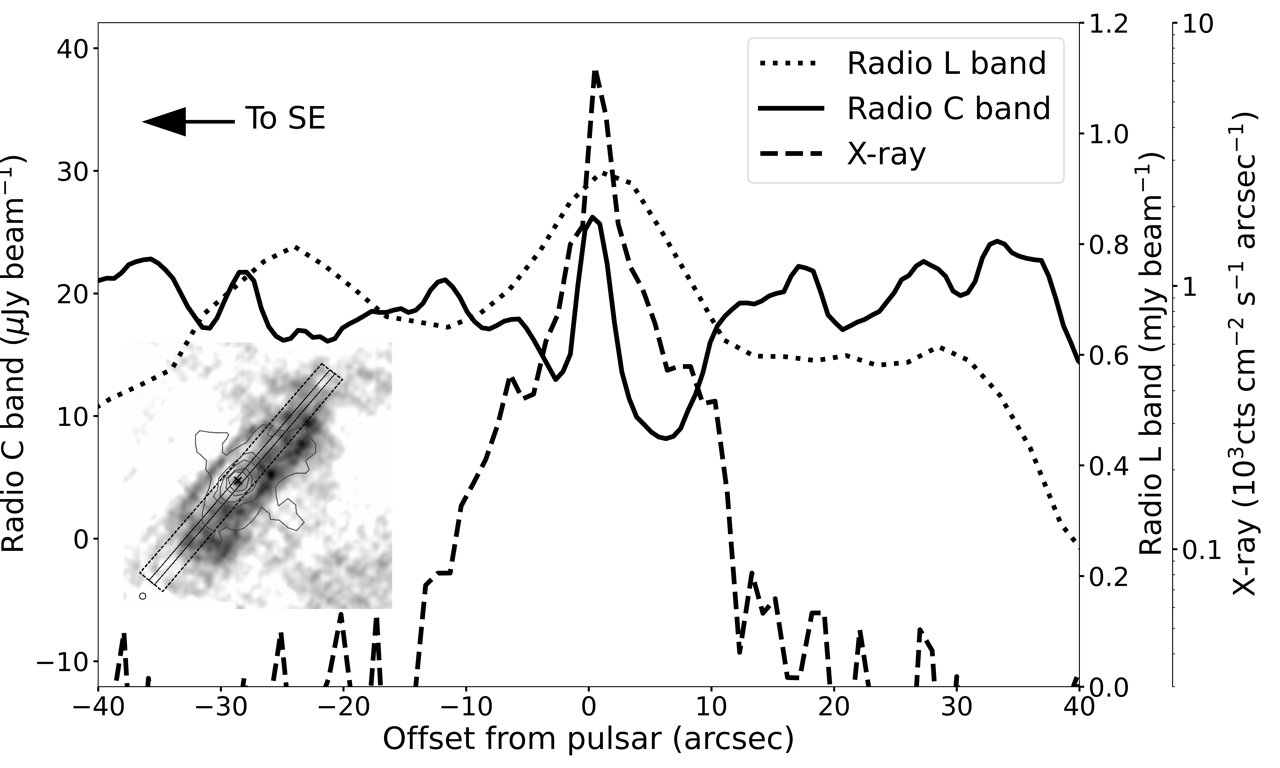}{1\columnwidth}{(b) Cross section along the major axis}}
\caption{Radio L-band (dotted), radio C-band (solid), and X-ray (dashed) brightness profiles along the (a) minor-axis and (b) the major axis cross-sections.
These are measured from a 12\arcsec\ wide region (the large rectangle in the inset plots) for the VLA L-band (1.5\,GHz) and a 3\arcsec\ wide region (the small rectangle in the inset plots) for both the VLA C-band (6\,GHz) and the \emph{Chandra} X-ray (0.3--7\,keV exposure-corrected and binned to 0.984$\arcsec$) intensity maps along two axes.}\label{fig:crosssection}
\end{figure}

We plot the cross-section profiles of the radio and X-ray intensity maps along the minor and major axes of the PWN in Figure~\ref{fig:crosssection}. It is clear that the central peak at the pulsar position is wider than the radio beam and the X-ray PSF, implying that there is diffused emission surrounding the pulsar in both radio and X-rays. This can be supported by the fact that the measured radio flux density at the pulsar location is $\sim$1\,mJy at the L-band, much larger than the reported pulsed emission of $\sim$0.1\,mJy \citep{Roberts2002}. The brightness profiles in all three bands are roughly asymmetric about the pulsar position and the intensity falls off more rapidly on the northeast (NE) side along the minor axis. 

Along the minor-axis cross-section, the C-band intensity drops from the central peak of $\sim28$\,mJy to $\sim$5\%--15\%\,mJy on either side of the pulsar in the $\sim$10\arcsec-wide ``valleys" and rises to a second peak (see Figure~\ref{fig:crosssection}a). The two second peaks correspond to bridge and the main-body region as seen in the radio C-band intensity map (see Figure~\ref{fig:radio_intensity}). The X-ray profile along the minor axis shows a flat bump (within $\sim5\arcsec$ west of the pulsar) on the south-west (SW)  side of the X-ray profile (Figure~\ref{fig:crosssection}a). It was fit with a double-torus model in previous X-ray studies \citep{Hessels2004, VanEtten2008}. In the L-band, the valleys and the second peaks are difficult to identify along the minor axis and are barely seen along the major axis probably due to resolution limit (FWHM $\sim$12\arcsec). The plateau emission profiles on the western side in both radio bands indicate that the nebula extends to the west, which agrees with the large scale X-ray emission detected with \emph{Chandra}  and \emph{XMM-Newton} \citep{VanEtten2008,Mizuno2017}. 

\subsection{Radio spectrum}\label{subsec:radio_spec}
\begin{figure}[htbp]
\plotone{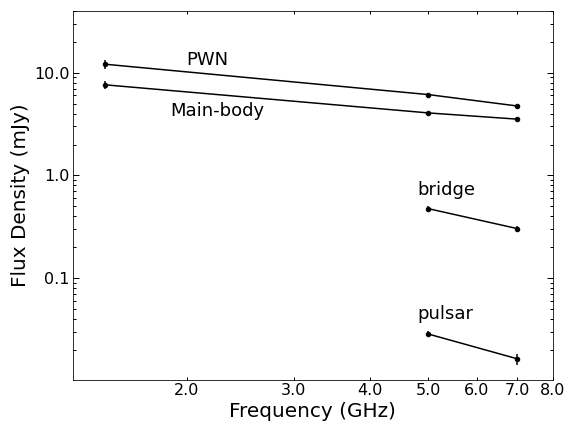}
\caption{Radio spectrum of the Dragonfly PWN. The flux densities at 1.5, 5, and 7\,GHz in the radio PWN, main-body, ``bridge", and in the pulsar regions. The solid lines denote the best-fit PL spectra. The resolution of the L-band is too low to resolve the bridge and pulsar regions. \label{fig:radio_spectrum}}
\end{figure}

We measure the C-band flux densities of the substructures as shown in Figure~\ref{fig:radio_intensity} at two sub-bands centered at 5 and 7\,GHz. For the L-band (1.47\,GHz), we only measure the flux densities of the larger regions due to limited resolution. The errors are estimated using the uncertainty of the calibration calibrator and the rms of the emission free region. The spectral results are plotted in Figure~\ref{fig:radio_spectrum}. 

The spectral index ($\alpha$) of different regions are obtained by fitting a power-law to the flux density (S$_{\nu}$) at different frequencies: $S_\nu\propto\nu^\alpha$. The results are listed in Table~\ref{tab:fluxspec}. The best-fit spectral index between 5 and 7\,GHz $\alpha_{\mathrm{5-7\,GHz}}$ is $-0.76\pm0.03$ for the overall radio PWN. The main-body region is much flatter (i.e., harder) ($\alpha_{\mathrm{5-7\,GHz}}=-0.41\pm0.04$) than the bridge region ($\alpha_{\mathrm{5-7\,GHz}}=-1.33\pm0.27$). If we take the 1.47\,GHz flux densities into account, the spectral index of the whole PWN and the main-body region between 1.47 and 7\,GHz are $\alpha_{\mathrm{1.5-7\,GHz}}=-0.74\pm0.02$ and $-0.44\pm0.04$, respectively, which are consistent with the result from only the C-band data. Since the angular size of the radio PWN is larger than the largest sensitive scale ($\sim35\arcsec$ at 5.0\,GHz and $\sim25\arcsec$ at 7.0\,GHz) of our observation, our data may not be able to recover all the flux densities at large scales. To check this, we also include the spectral index fitting between 1.47\,GHz and 5\,GHz. The resulting spectral indexes are flatter ($\alpha_{\mathrm{1.5-5\,GHz}}=-0.62\pm0.08$) for the whole PWN but steeper ($\alpha_{\mathrm{1.5-5\,GHz}}=-0.51\pm0.07$) for the main-body region than those of either 1.5$-$7\,GHz or 5$-$7\,GHz. Therefore we argue that this is not an issue for the main-body, since the two spectral indexes $\alpha_{\rm 1.5-7\,GHz}$ and $\alpha_{\rm 5-7\,GHz}$ are consistent.

\begin{deluxetable}{lccc}
\tablecaption{Radio Spectral Indexes of the Different Parts of the Dragonfly PWN\label{tab:fluxspec}}
\tablehead{
\colhead{Region Name} & \colhead{$\alpha_{\rm 1.5-7\,GHz}$} & \colhead{$\alpha_{\rm 1.5-5\,GHz}$} &\colhead{$\alpha_{\rm 5-7\,GHz}$}}
\startdata
PWN & $-0.74\pm0.02$ & $-0.62\pm0.08$ & $-0.76\pm0.03$ \\
Main-body & $-0.44\pm0.0$ & $-0.51\pm0.07$ & $-0.41\pm0.04$ \\
Bridge &  \nodata  &  \nodata & $-1.33\pm0.27$ \\
Pulsar &  \nodata  &  \nodata & $-1.65\pm0.42$
\enddata
\tablecomments{The regions are shown in Figure~\ref{fig:radio_intensity}.}
\end{deluxetable}

\begin{figure}[htbp]
\includegraphics[width=1.0\columnwidth]{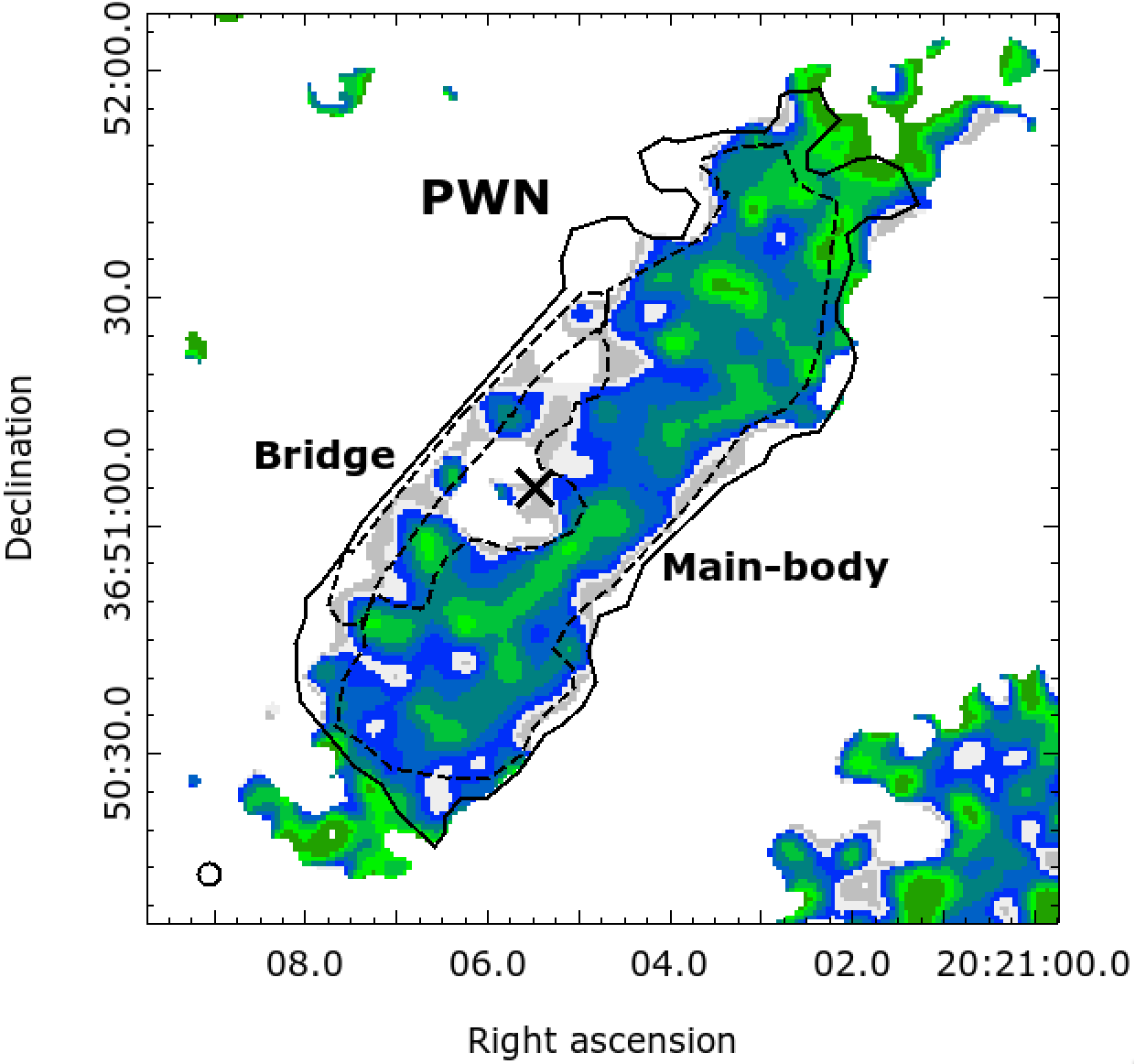}
\caption{Spectral index map of the Dragonfly PWN between 5 and 7\,GHz. The map is clipped if the total intensity in either map is below 1-$\sigma$ detection. The circle located at the lower left corner shows the restored beam size and the location of PSR J2021+3651 is marked with an ``x". The PWN region is the same as that in Figure~\ref{fig:radio_intensity}.\label{fig:alpha}}
\end{figure}

Finally, we calculate the spectral index map between the 5\,GHz and 7\,GHz images and the result is shown in Figure~\ref{fig:alpha}. The spectral index increases (flattens) gradually from the bridge with the $\alpha\lesssim-1$ to the main-body region with the mean and median spectral indexes of $-0.1$ and 0.1 respectively.

\subsection{Multi-wavelength spectrum}\label{subsec:multi_spec}
\begin{figure}
\centering
\plotone{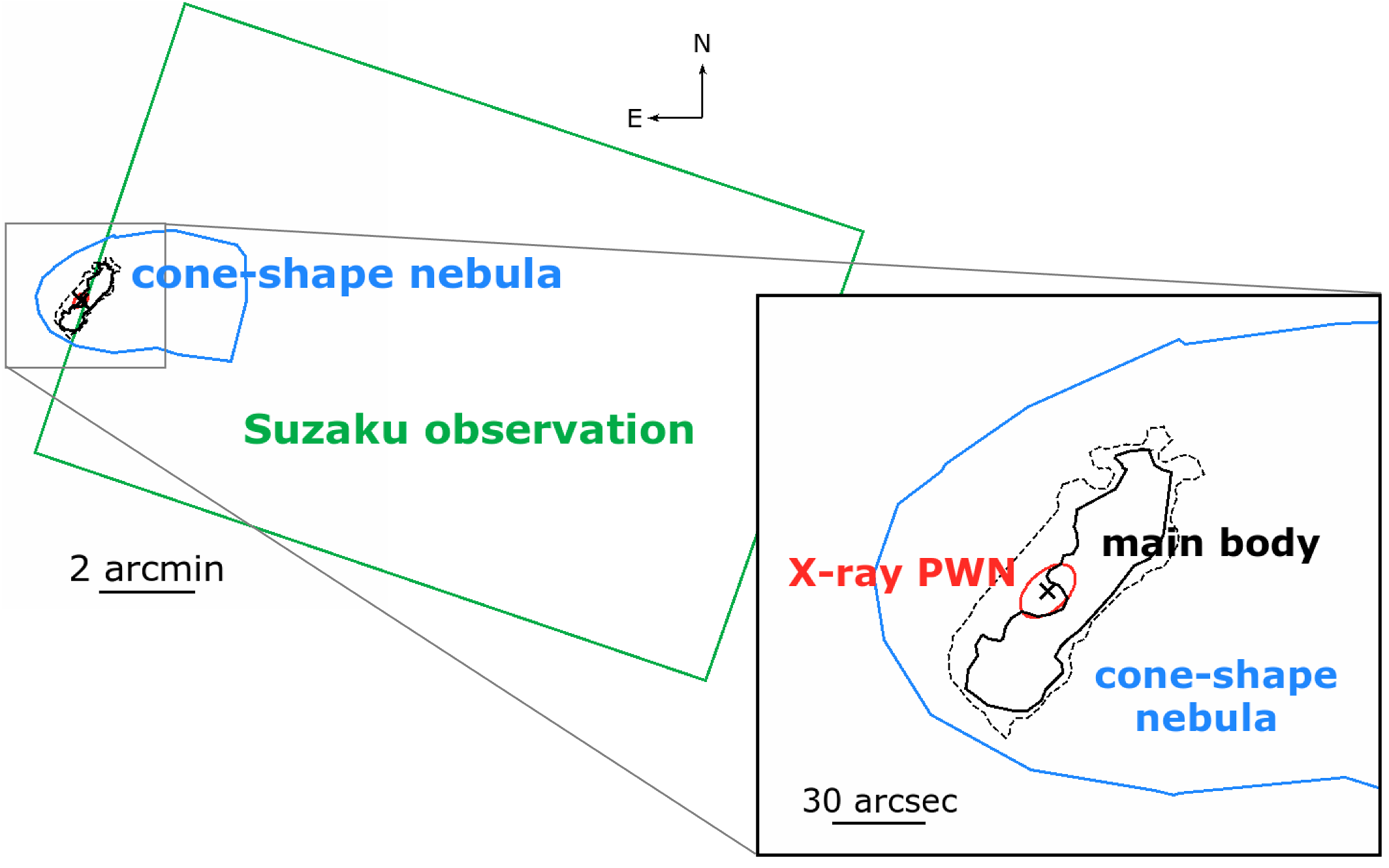}
\plotone{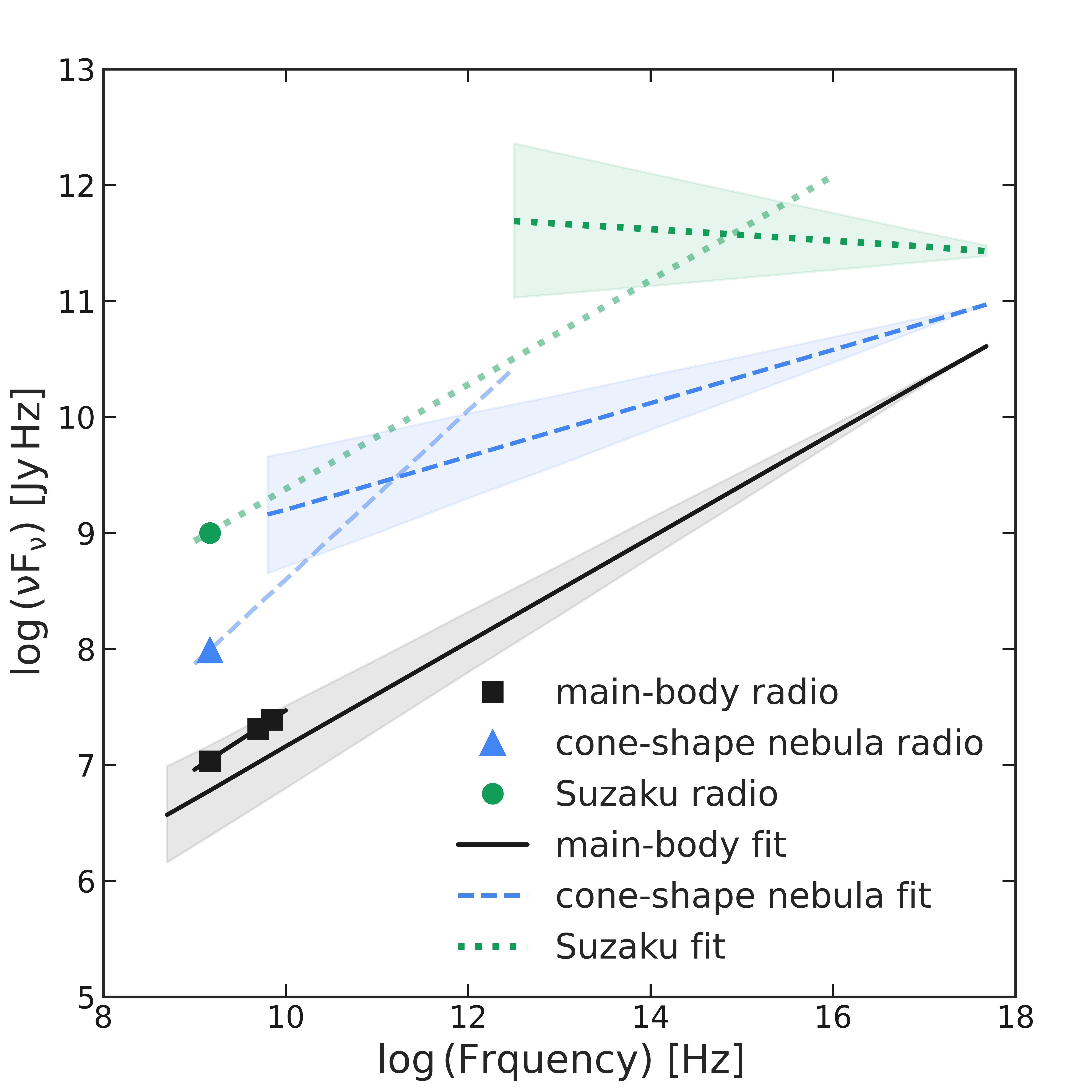}
\caption{Top: extraction regions for the X-ray spectra of the Dragonfly. From largest to smallest: the \emph{Suzaku} observation region \citep{Mizuno2017} (green), the ``cone-shape nebula" region (blue), the C-band main-body (black), and the X-ray PWN \citep{Kirichenko2015} (red). Bottom: SEDs of different regions in the top figure. The solid line indicates the main-body spectrum. For comparison, spectra from the cone-shape nebula region (excluding the X-ray PWN region), which similar to the ``outer nebula'' region in \citet{VanEtten2008}, and the \emph{Suzaku} observation region from \citet{Mizuno2017} are shown by the dashed and dotted lines, respectively. We extrapolated the lines to the radio band by assuming a spectral break with $\Delta\alpha=0.5$ due to synchrotron cooling.}\label{fig:multi_spectrum} \end{figure}

Our X-ray spectral analysis found a photon index of $\Gamma=1.55\pm0.05$ for the main body, slightly softer than $\Gamma=1.47\pm0.04$ for the central X-ray PWN (excluding the inner radius of 2\arcsec\ pulsar region). 
Moving further out, we found that the X-ray spectrum softens to $\Gamma=1.77\pm0.06$ in the cone-shape nebula region (excluding the main-body region) and $\Gamma=2.05\pm0.12$ in an even larger region (15$\arcmin\times10\arcmin$) based on \emph{Suzaku} observation \citep{Mizuno2017}. The resulting broadband spectral energy distributions (SEDs) are plotted in Figure~\ref{fig:multi_spectrum}.
We find that the X-ray and radio spectra of the main-body region can be connected with a single, unbroken power law. For the cone-shape nebula and the \emph{Suzaku} observation regions, the C-band data do not have enough sensitivity to provide flux density measurements. We therefore show in the plot only the L-band data and assumed a spectral break with $\Delta \alpha=0.5$ between radio and X-rays, as expected from synchrotron cooling \citep[see][for details]{Pacholczyk1970}. 

\subsection{Polarization} \label{subsec:result_bfield}
\begin{figure*}[htb]
\includegraphics[width=1\textwidth]{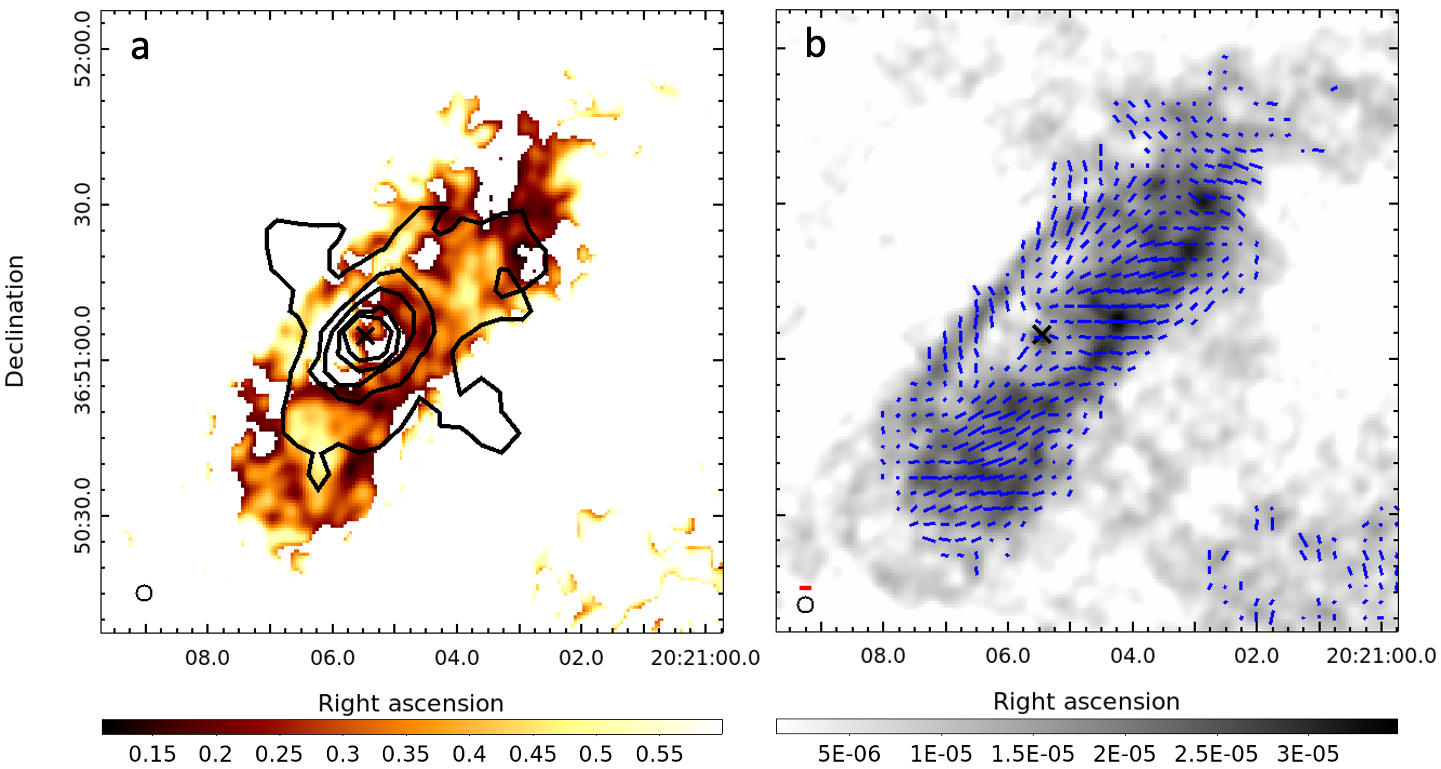}
\caption{(a) Fractional linearly polarization of the Dragonfly PWN at 6\,GHz, overlaid with the \emph{Chandra} 0.3--7\,keV X-ray contours. The radio beam size is shown in lower left. (b) Intrinsic $B$-field orientation of the Dragonfly PWN overlaid on the total intensity map. The vector lengths are proportional to the polarized intensity.
The red scale bar at the lower left represents linear polarization intensity of 10\,$\mu$Jy\,beam$^{-1}$.
Both maps are clipped if the polarization intensity has S/N $<3$.
\label{fig:polarization}}
\end{figure*}

Figure~\ref{fig:polarization}a shows the fractional polarization map. The entire PWN is highly linearly polarized with polarization fraction (PF) $\gtrsim30\%$ and even $>40$\% in the two wings of the main-body and the bridge regions. The PF distribution is non-uniform and appears to trace the X-ray PWN morphology, so it is unlikely due to noise. The PF is higher on the X-ray PWN wings but much lower ($\leq30\%$) in the central part.

We attempted to derive a rotation measure (RM) map using the polarization angles in the two subbands at 5\,GHz and 7\,GHz. However, the resulting RM map has too large uncertainty to be useful. We therefore adopt a constant RM value of $-524$\,rad\,m$^{-2}$ of the pulsar \citep{Abdo2009} to derotate the polarization vectors. The intrinsic magnetic field orientation after Faraday rotation correction is shown in Figure~\ref{fig:polarization}b.
The magnetic field in the bridge region is predominately along the north-south direction. It then switches over the gap region to mostly east-west in the main-body. The change in direction is unlikely caused by fluctuation of the foreground RM as it shows a good correlation with the overall PWN structure. Finally, we note that there are polarization signals detected in the outer lobes south-west of the PWN.  

\section{Discussion}\label{sec:discussion}
\subsection{Pulsar motion}
The radio band emission of the Dragonfly in the L-band shows an overall bow-shock morphology, suggesting supersonic motion of the pulsar in the ISM. This is supported by the asymmetry between the NW and SE wings seen in the radio C-band and X-rays, and the distortion of the X-ray jet and the $B$-field vectors. Nonetheless, we note that the torus-jet structure and the toroidal $B$-field geometry are not totally disrupted. These imply that the pulsar motion is at most mildly supersonically. For a typical supersonically moving pulsar, its ram pressure is so large that its microstructures would be destroyed \citep[see examples in ][]{Kargaltsev2017}.

The apex of the bow shock structure in both radio L-band and X-rays is $\sim$50\arcsec\ from the pulsar, corresponding to a physical scale $r\approx0.4$\,pc at 1.8\,kpc. If we take this as the bow shock stand-off distance, the pulsar velocity ($v_{\rm PSR}$) can be estimated from the balance between the PWN internal pressure and the ram pressure, such that $r=\sqrt{\dot E /4\pi c\rho v_{\rm PSR}^2}$, where $\rho$ is the ambient density. The latter is difficult to determine and the typical values in the cold and hot phases of the ISM give a wide range of $v_{\rm PSR}$ estimates from a few to hundreds of km per second  \citep{Barkov2019,VanEtten2012}.

If the pulsar was born at the centroid of the TeV emission, the $\sim$0.3\arcdeg\ offset suggests a high transverse velocity of $\sim$300--2500\,km\,s$^{-1}$ at a distance of $1.8_{-1.4}^{+1.7}$\,kpc, where the large range of the velocity estimate mostly come from the distance uncertainty. This is possible but the upper bound lies in the high end of the pulsar velocity distribution \citep{Kargaltsev2017}. In any case, the velocity is clearly supersonic given that the typical ISM sound speed is $\sim$10--100\,km\,s$^{-1}$. We note that the proper motion inferred by the TeV position misaligns with the pulsar spin axis. It would affect the $B$-field orientation and its interaction with ambient ISM according to the 3-D simulations \citep{Barkov2019,Barkov2019jet} that will be discussed in the later subscetions.

\subsection{Radio PWN Structure} \label{subsec:discuss_morph}

At a distance of 1.8\,kpc, the elliptical morphology of the Dragonfly PWN in radio corresponds to a physical size of 0.9\,pc$\times$0.3\,pc, comparable to the Vela PWN in radio \citep{Dodson2003}. The radio emission of the Dragonfly has the same overall orientation as the X-ray counterpart, but about two times as large in size. This is a common feature among PWNe due to the longer synchrotron cooling time of the radio-emitting particles.

We can compare the Dragonfly with Vela and B1706 as their central pulsars share very similar properties, including characteristic age, spin-down power, and $\gamma$-ray pulsations. On the other hand, their X-ray PWN structures such as torus and jet show different morphology \citep{Dodson2003, 2004IAUS..218..195K,Romani2005,Liu2022}. The radio emission of Vela and B1706 PWNe are found outside the X-ray regions and it is faint at the inner PWN. However, the two wings for the Dragonfly in radio are connected and overlap with the X-ray emission. It is unclear if the difference is due to different viewing geometry or other physical mechanisms.   

In the L-band image, a protruding structure ahead of bow shock (see Figure~\ref{fig:zoomout}) was also found in the \emph{Chandra} image \citep{VanEtten2008}. The 3-D simulation suggested these kinetic jets are formed due to the leaking PWN particles into the ISM \citep{Barkov2019, Barkov2019jet}. These particles escape along the reconnected magnetic field between the PWN and the ISM \citep{Bandiera2008}. 

Moving to small scale, our high resolution radio map reveals, for the first time, peculiar features in the Dragonfly. The most intriguing one is the narrow bridge region that has different properties than the main body, including steeper radio spectrum, different $B$-field orientation, and no X-ray counterpart. It was suggested that the pulsar was born near the TeV source and moves towards east. In this picture, the bridge is then the leading edge of the PWN compressed by the ISM ram pressure. The compression could enhance the $B$-field and hence the synchrotron emissivity, and perhaps increases the polarization fraction. This is supported by the sharp edge of the bridge in NE in both the C- and L-band images in Figure~\ref{fig:crosssection}a and also slight deformation of the X-ray PWN in NE as indicated by the white X-ray contours in Figure~\ref{fig:zoomin_3color}. The pulsar motion could also make the NW wing, which trails the pulsar motion, longer than the SE wing of the main body, in both radio and X-ray bands.

Alternatively, the observed bridge could correspond to the far side of the radio PWN if it has a ring-like structure in 3-D and is viewed near edge-on. Such a ring (with a diameter of $\sim$0.6\,pc in the Dragonfly) could resemble that formed at the termination shock detected in the Crab Nebula in X-rays and radio \citep[e.g.,][]{Krassilchtchikov2017, Dubner2017} and the putative ring seen in the Vela PWN \citep{Ponomaryov2018}. A ring-like structure is observed in X-rays \citep[called 'double tori' in][]{VanEtten2008} but is smaller than the radio ring. Perhaps they are formed by different populations of particles. Another possibility is that the X-ray ring is newly produced, just like the X-ray ring of the young Crab Nebula, while the radio ring is the remaining older ring with its X-ray component having cooled down. In this case, the bridge would appear as fainter and narrower than the main-body, which corresponds to the near side, due to Doppler boosting \citep[see][]{Ng2004, NgRomani2008}. The brightness ratio between the two regions is $R\approx1.15$. The inclination angle $\zeta$ between the ring axis and the line of sight can be deduced by the relation $R=[(1+\beta\cos\zeta)/(1-\beta\cos\zeta)]^{2-\alpha}$. We adopt the best-fit post-shock velocity $\beta\sim0.8$ inferred by the X-ray torus fitting \citep{VanEtten2008} and the C-band spectral index $\alpha=-0.4$ of the main-body. These give $\zeta\approx88\arcdeg$, which is very close to $\sim85\arcdeg$ from fitting the X-ray torus \cite{VanEtten2008}. This can also explain the rapid fall off of the L-band and C-band intensity profiles seen on the eastern side of the minor-axis cross section in Figure~\ref{fig:crosssection}a. This model, however, is less preferred, as the ring size is much larger than the X-ray tori. The latter is believed to correspond to the wind termination shock. The formation mechanism of such a radio ring is unclear. This picture also has difficulty explaining the different spectral index in the bridge and the main-body. 

\subsection{Multi-wavelength Spectral Analysis}\label{subsec:discuss_spec}
The radio spectral index of the overall radio PWN is $\sim-0.7$ to $-0.8$, which is much steeper, i.e.\ softer, than the typical radio PWN spectrum ($-0.3\lesssim\alpha\lesssim0$), such as Crab ($\alpha=-0.27$ \citealp{Strom&Greidanus1992}) and G283.2--0.59 ($\alpha=-0.13$ \citealp{Ng2017}). As we mentioned, this soft emission is mainly contributed by the bridge region, probably due to compression as the pulsar moves in the ISM. If we consider only the main-body region, which has larger S/N and contributes $\sim$70\% of the PWN flux density at 6\,GHz (see Figure~\ref{fig:radio_spectrum}), its spectral index is $\sim-0.4$, more inline with the typical value. 

The SED of the main-body region from radio to X-rays can be fit with a single, unbroken powerlaw as shown in Figure~\ref{fig:multi_spectrum}. This suggests no significant synchrotron burn-off, implying a young age for the particles. For the cone-shape nebula and the \emph{Suzaku} region, if we assume a break with $\Delta\alpha=0.5$ between the X-ray and radio spectra, we can infer spectral turn-over ($\nu_{\rm b}$) at $\log [\nu_{\rm b}(\mbox{Hz})]=11.2_{-1.0}^{+0.7}$ and $14.9_{-1.0}^{+0.5}$, respectively. Future direct measurement of their radio spectra can confirm this. 

\subsection{Nebular Magnetic Field Strength}\label{subsec:discuss_bstrength}
We estimate the so-called equipartition magnetic field strength ($B_{\rm eq}$) of the PWN by minimizing the total energy of the particles and the magnetic field. Using standard synchrotrn theory, it can be shown that $B_{\rm eq}$ is related to the synchrotron luminosity $L$ and the emission volume $V$ by
\begin{equation}\label{eqn:equipartition}
B_{\rm eq}=\left (6 \pi c_{12}\frac{L}{V}\right )^{2/7},  
\end{equation}
where $c_{12}$ is a constant weakly dependent on the spectral index and the frequency range \citep{Pacholczyk1970}. Assuming that the overall PWN has a cone structure in 3D as suggested by the L-band image and taking the main body and the bridge regions as slices of the cone, we estimate their volumes of $\sim4.1\times10^{54}\,d^3_{1.8}$\,cm$^3$ and $\sim3.9\times10^{53}\,d^3_{1.8}$\,cm$^3$, respectively, where $d_{1.8}$ is the source distance in units of 1.8\,kpc. The calculated luminosities ($L$) are 6.9$\times10^{29}\,d^2_{1.8}$erg\,s$^{-1}$ for the main-body and 2.3$\times10^{29}\,d^2_{1.8}$erg\,s$^{-1}$ for the bridge region, based on the integrated flux in the frequency range of $10^7-10^{11}$\,Hz. These give $B_{\rm eq}$ of $\sim22$\,$\mu$G for the main-body and a higher value of $\sim59$\,$\mu$G for the bridge. The result could indicate magnetic field enhancement due to compression by the ISM ram pressure.

For the cone-shape nebula region excluding the inner PWN, similar estimate gives $B_{\rm eq}\approx 21\,\mu$G assuming the spectrum shown in Figure~\ref{fig:multi_spectrum}. This is very close to the estimate of $\sim20\,\mu$G by extrapolating the $B$-field from the pulsar surface \citep{VanEtten2008}. Similarly, by taking the \emph{Suzaku} observation region as a cylinder in 3D, we derived $B_{\rm eq}\approx19\,\mu$G, inline with the value for the cone-shape nebula above. 

Synchrotron cooling can give rise to a spectral break at frequency
\begin{equation}\label{eqn:break}
\nu_{b}=10^{21}\left(\frac{B}{10^{-6}\mathrm{G}}\right)^{-3}\left(\frac{t}{10^3\mathrm{yr}}\right)^{-2}\,\mathrm{Hz}, 
\end{equation}
where $B$ is the magnetic field strength and $t$ is the age of the system. Using $B\sim21\,\mu$G estimated above and the pulsar's age of 7\,kyr, we expect $\nu_{\rm b}\sim10^{15}$\,Hz in the cone-shape nebula. This is much higher than the spectral turn-over at $\nu_{\rm b}\sim10^{11}$\,Hz inferred from a break of $\Delta\alpha=0.5$ between the X-ray and radio spectra (see Figure~\ref{fig:multi_spectrum}). This can be reconciled if the actual $B$-field is higher than the equipartition estimate or the true age is much larger than 7\,kyr. The value $\nu_{\rm b}\sim10^{15}$\,Hz obtained from the equation above, on the other hand, is consistent with the inferred cooling break for the \emph{Suzaku} observation region. X-ray spectral analysis found no obvious photon index variation in the \emph{Suzaku} region \citep{Mizuno2017}, suggesting that the particles probably have already cooled down and hence the inferred spectral break in Figure~\ref{fig:multi_spectrum} could be real.

\subsection{Magnetic Field Geometry}\label{subsec:discuss_bfield}

Our high resolution polarization maps reveal different $B$-field configurations in different part of the Dragonfly: it is oriented along the north-south direction in the bridge and east-west in the main body. There is no obvious connection with the X-ray torus structure, contrary to a toroidal $B$-field expected in young PWNe \citep[e.g.,][]{Porth2017}. On the other hand, the field geometry seems better correlate with the inferred pulsar motion direction, similar to some bow-shock PWNe, e.g., the Mouse (G359.23$-$0.82) and the Frying Pan (G315.78$-$0.23) \citep{Yusef2005,ng2012}. In particular, the Mouse has a $B$-field perpendicular to the pulsar motion ahead of the bow-shock apex. It then switches to parallel after a depolarized region at the shock \citep{Yusef2005}. For the Dragonfly, the $B$-field also changes direction across the faint gap between the bridge and the main body. Note that the Mouse is believed to be moving highly supersonically \citep[see][]{Klingler2018} but the Dragonfly is at most mildly supersonically as we argued. It is intriguing that they share some similarities in the $B$-field geometry. This suggests that the Mach number of the pulsar motion is not the only factor determining the field configuration.

In the main body, the magnetic field orientation can be described by the sum of two components: one follows the torus direction and one follows the pulsar moving direction. This is different than the Mouse, which has $B$-field parallel to the pulsar tail. It was suggested that the central pulsar of the Mouse has spin axis perpendicular to its motion \citep{Klingler2018}, such that the two $B$-field components align. 

On the other hand, the magnetic field configuration in the bridge region is different from that of the main body. It fits better with the picture of $B$-field reconnection between the pulsar wind and the ISM in the head of the PWN, as supported by the 3-D simulation \citep{Barkov2019, Barkov2019jet}. This $B$-field reconnection allow relativistic particles escaping from the pulsar wind and form the kinetic jet as discussed in Section~\ref{subsec:discuss_morph}.

Finally, we note that our polarization result is inconsistent with the idea that the bridge is the far side of a ring-like structure in 3D, since that should give a similar $B$-field geometry in both the bridge and the main body.

\subsection{Diffuse Emission Lobes}
There are a few diffuse emission lobes located southwest of the pulsar (see Figure~\ref{fig:zoomout}a). These could be larger scale emission resolved out due to lack of short spacing in our observation. We argue that they are likely real emission rather than sidelobes, as they are located in the pulsar's path and there is also patchy radio emission detected in the L-band. They could be relic pulsar wind particles interacting with inhomogeneous ISM \citep{Toropina2019} or leftover after the pulsar passed by.

For the southernmost lobe shown in the C-band image in Figure~\ref{fig:zoomout}a, its emission in the L-band is rather faint. If this extended emission is real, it could indicate an inverted spectrum, different than that of the inner PWN. This requires some re-acceleration mechanism, or it could be caused by instability at the bow-shock interface. 

\section{Conclusion}\label{sec:conclusion}

In this paper, we reported on a detailed radio study of the Dragonfly PWN using new VLA observations at the C-band (4--8\,GHz) together with archival radio L-band VLA data and \emph{Chandra} X-ray data. The radio emission in the C-band shows elongated structure, with size twice as large as the X-ray counterpart. It consists of a bright, axisymmetric main-body in the west and a narrow bridge in the east. We suggest that the latter is due to compression by the ram pressure when the pulsar travels supersonically in the ISM. This is supported by the high polarization fraction, the soft spectrum ($\alpha_{\rm{1.5-7\,GHz}}=-1.3$) and the $B$-field configuration of the bridge. For the main-body region, the radio spectrum is harder ($\alpha_{\rm{1.5-7\,GHz}}=-0.4$), closer to that of a typical PWN, and it can be connected to the X-ray spectrum by a power law without a break. This suggests that the X-ray emitting particles are probably not yet cooled down.

The radio emission of the Dragonfly at 6\,GHz has a high linear polarization fraction of over 30\% in most regions and even higher in the two wings of the X-ray PWN and the bridge. This indicates a highly ordered magnetic field structure. The intrinsic $B$-field in the bridge predominately are perpendicular to the pulsar motion. It then switches direction across the faint gap and the field configuration in the main body can be described by a toroidal pattern slightly distorted by the pulsar motion. Similar behavior was observed in another PWN that travels highly supersonically in the ISM. However, for the case of the Dragonfly, we argue that the pulsar moves transonically or mildly supersonically since the stand-off distance is large, its jet and torus structures and the $B$-field are not disrupted. 
    
\begin{acknowledgments}

Ruolan Jin is supported by the Yushan Fellow Program of the Ministry of Education of the Republic of China (Taiwan). C.-Y. Ng is supported by a GRF grant of the Hong Kong Government under HKU 17301618. K.L.L. is supported by the Ministry of Science and Technology of the Republic of China (Taiwan) through grant 110-2636-M-006-01 and is a Yushan Young Fellow of the Ministry of Education of the Republic of China (Taiwan). The National Radio Astronomy Observatory is a facility of the National Science Foundation operated under cooperative agreement by Associated Universities, Inc. This research has made use of data obtained from the Chandra Data Archive and software provided by the Chandra X-ray Center (CXC) in the application packages CIAO and Sherpa.

\end{acknowledgments}
%

\vspace{5mm}
\facilities{VLA, CXO(ACIS)}


\software{CASA \citep[5.6.1-8]{2007ASPC..376..127M}, 
CIAO, MIRIAD \citep{Sault1995}}, 
CIAO \citep{Fruscione2006}, 
Shepra \citep{Freeman2001}

\bibliography{jinreference}

\begin{thebibliography}{}
\expandafter\ifx\csname natexlab\endcsname\relax\def\natexlab#1{#1}\fi
\providecommand{\url}[1]{\href{#1}{#1}}
\providecommand{\dodoi}[1]{doi:~\href{http://doi.org/#1}{\nolinkurl{#1}}}
\providecommand{\doeprint}[1]{\href{http://ascl.net/#1}{\nolinkurl{http://ascl.net/#1}}}
\providecommand{\doarXiv}[1]{\href{https://arxiv.org/abs/#1}{\nolinkurl{https://arxiv.org/abs/#1}}}

\bibitem[{{Abdo} {et~al.}(2009){Abdo}, {Ackermann}, {Ajello}, {Atwood},
  {Baldini}, {Ballet}, {Barbiellini}, {Bastieri}, {Battelino}, {Baughman},
  {Bechtol}, {Bellazzini}, {Berenji}, {Bloom}, {Bogaert}, {Borgland},
  {Bregeon}, {Brez}, {Brigida}, {Bruel}, {Burnett}, {Caliandro}, {Cameron},
  {Camilo}, {Caraveo}, {Casandjian}, {Cecchi}, {Charles}, {Chekhtman}, {Chen},
  {Cheung}, {Chiang}, {Ciprini}, {Cognard}, {Cohen-Tanugi}, {Cominsky},
  {Conrad}, {Cutini}, {Demorest}, {Dermer}, {de Angelis}, {de Luca}, {de
  Palma}, {Digel}, {Dormody}, {do Couto e Silva}, {Drell}, {Dubois}, {Dumora},
  {Espinoza}, {Farnier}, {Favuzzi}, {Focke}, {Frailis}, {Freire}, {Fukazawa},
  {Funk}, {Fusco}, {Gargano}, {Gasparrini}, {Gehrels}, {Germani}, {Giebels},
  {Giglietto}, {Giordano}, {Glanzman}, {Godfrey}, {Grenier}, {Grondin},
  {Grove}, {Guillemot}, {Guiriec}, {Hanabata}, {Harding}, {Hayashida}, {Hays},
  {Hughes}, {J{\'o}hannesson}, {Johnson}, {Johnson}, {Johnson}, {Johnson},
  {Johnston}, {Kamae}, {Katagiri}, {Kataoka}, {Kawai}, {Kerr}, {Kiziltan},
  {Kn{\"o}dlseder}, {Komin}, {Kramer}, {Kuehn}, {Kuss}, {Lande}, {Latronico},
  {Lee}, {Lemoine-Goumard}, {Longo}, {Loparco}, {Lott}, {Lovellette},
  {Lubrano}, {Lyne}, {Makeev}, {Manchester}, {Marelli}, {Mazziotta},
  {McConville}, {McEnery}, {McLaughlin}, {Meurer}, {Michelson}, {Mitthumsiri},
  {Mizuno}, {Moiseev}, {Monte}, {Monzani}, {Morselli}, {Moskalenko}, {Murgia},
  {Nolan}, {Noutsos}, {Nuss}, {Ohsugi}, {Omodei}, {Orlando}, {Ormes}, {Ozaki},
  {Paneque}, {Panetta}, {Parent}, {Pepe}, {Pesce-Rollins}, {Piron}, {Porter},
  {Rain{\`o}}, {Rando}, {Ransom}, {Razzano}, {Reimer}, {Reimer}, {Reposeur},
  {Ritz}, {Rochester}, {Rodriguez}, {Romani}, {Ryde}, {Sadrozinski}, {Sanchez},
  {Parkinson}, {Sgr{\`o}}, {Sierpowska-Bartosik}, {Siskind}, {Smith}, {Smith},
  {Spandre}, {Spinelli}, {Stappers}, {Starck}, {Strickman}, {Suson},
  {Tajima}https://www.overleaf.com/project/5ed646b0b2d0a60001a50d55,
  {Takahashi}, {Takahashi}, {Tanaka}, {Thayer}, {Thayer}, {Theureau},
  {Thompson}, {Thorsett}, {Tibaldo}, {Torres}, {Tosti}, {Tramacere},
  {Uchiyama}, {Usher}, {Van Etten}, {Vilchez}, {Vitale}, {Waite}, {Wallace},
  {Watters}, {Weltevrede}, {Wood}, {Ylinen}, \& {Ziegler}}]{Abdo2009}
{Abdo}, A.~A., {Ackermann}, M., {Ajello}, M., {et~al.} 2009, \apj, 700, 1059,
  \dodoi{10.1088/0004-637X/700/2/1059}

\bibitem[{{Albert} {et~al.}(2021){Albert}, {Alfaro}, {Alvarez},
  {Arteaga-Vel{\'a}zquez}, {Arunbabu}, {Avila Rojas}, {Ayala Solares},
  {Baghmanyan}, {Belmont-Moreno}, {Brisbois}, {Caballero-Mora},
  {Capistr{\'a}n}, {Carrami{\~n}ana}, {Casanova}, {Cotzomi}, {Couti{\~n}o de
  Le{\'o}n}, {De la Fuente}, {Diaz Hernandez}, {Dingus}, {DuVernois},
  {Durocher}, {Engel}, {Espinoza}, {Fraija}, {Garcia},
  {Garc{\'\i}a-Gonz{\'a}lez}, {Giacinti}, {Gonz{\'a}lez}, {Goodman}, {Harding},
  {Hinton}, {Hona}, {Huang}, {Hueyotl-Zahuantitla}, {Huentemeyer},
  {Jardin-Blicq}, {Joshi}, {Lee}, {Le{\'o}n Vargas}, {Linnemann}, {Longinotti},
  {Luis-Raya}, {Lundeen}, {L{\'o}pez-Coto}, {Malone}, {Martinez},
  {Mart{\'\i}nez-Castro}, {Matthews}, {Miranda-Romagnoli}, {Morales-Soto},
  {Moreno}, {Mostaf{\'a}}, {Nayerhoda}, {Nellen}, {Newbold}, {Nisa},
  {Noriega-Papaqui}, {Olivera-Nieto}, {Omodei}, {Peisker}, {P{\'e}rez Araujo},
  {P{\'e}rez-P{\'e}rez}, {Rho}, {Rosa-Gonz{\'a}lez}, {Ruiz-Velasco}, {Salazar},
  {Salesa Greus}, {Sandoval}, {Schneider}, {Schoorlemmer}, {Serna-Franco},
  {Smith}, {Springer}, {Surajbali}, {Tollefson}, {Torres}, {Turner},
  {Ure{\~n}a-Mena}, {Weisgarber}, {Willox}, {Zhou}, {de Le{\'o}n}, \& {HAWC
  Collaboration}}]{Albert2021}
{Albert}, A., {Alfaro}, R., {Alvarez}, C., {et~al.} 2021, \apj, 911, 143,
  \dodoi{10.3847/1538-4357/abecda}

\bibitem[{{Aliu} {et~al.}(2014){Aliu}, {Aune}, {Behera}, {Beilicke}, {Benbow},
  {Berger}, {Bird}, {Bouvier}, {Buckley}, {Bugaev}, {Cerruti}, {Chen},
  {Ciupik}, {Connolly}, {Cui}, {Dumm}, {Dwarkadas}, {Errando}, {Falcone},
  {Federici}, {Feng}, {Finley}, {Fleischhack}, {Fortin}, {Fortson}, {Furniss},
  {Galante}, {Gillanders}, {Gotthelf}, {Griffin}, {Griffiths}, {Grube}, {Gyuk},
  {Hanna}, {Holder}, {Hughes}, {Humensky}, {Johnson}, {Kaaret}, {Kargaltsev},
  {Kertzman}, {Khassen}, {Kieda}, {Krennrich}, {Lang}, {Madhavan}, {Maier},
  {McArthur}, {McCann}, {Millis}, {Moriarty}, {Mukherjee}, {Nieto},
  {O'Faol{\'a}in de Bhr{\'o}ithe}, {Ong}, {Otte}, {Pandel}, {Park}, {Pohl},
  {Popkow}, {Prokoph}, {Quinn}, {Ragan}, {Rajotte}, {Reyes}, {Reynolds},
  {Richards}, {Roache}, {Roberts}, {Sembroski}, {Shahinyan}, {Smith},
  {Staszak}, {Telezhinsky}, {Tucci}, {Tyler}, {Vincent}, {Wakely}, {Weinstein},
  {Welsing}, {Wilhelm}, {Williams}, \& {Zitzer}}]{Aliu2014}
{Aliu}, E., {Aune}, T., {Behera}, B., {et~al.} 2014, \apj, 788, 78,
  \dodoi{10.1088/0004-637X/788/1/78}

\bibitem[{{Bandiera}(2008)}]{Bandiera2008}
{Bandiera}, R. 2008, \aap, 490, L3, \dodoi{10.1051/0004-6361:200810666}

\bibitem[{{Barkov} {et~al.}(2019){Barkov}, {Lyutikov}, \&
  {Khangulyan}}]{Barkov2019}
{Barkov}, M.~V., {Lyutikov}, M., \& {Khangulyan}, D. 2019, \mnras, 484, 4760,
  \dodoi{10.1093/mnras/stz213}

\bibitem[{Barkov {et~al.}(2019)Barkov, Lyutikov, Klingler, \&
  Bordas}]{Barkov2019jet}
Barkov, M.~V., Lyutikov, M., Klingler, N., \& Bordas, P. 2019, Monthly Notices
  of the Royal Astronomical Society, 485, 2041, \dodoi{10.1093/mnras/stz521}

\bibitem[{{Blasi} \& {Amato}(2011)}]{Blasi2011}
{Blasi}, P., \& {Amato}, E. 2011, Astrophysics and Space Science Proceedings,
  21, 624.
\newblock \doarXiv{1007.4745}

\bibitem[{{Briggs}(1995)}]{Briggs1995}
{Briggs}, D.~S. 1995, in American Astronomical Society Meeting Abstracts, Vol.
  187, American Astronomical Society Meeting Abstracts, 112.02

\bibitem[{{Della Torre} {et~al.}(2015){Della Torre}, {Gervasi}, {Rancoita},
  {Rozza}, \& {Treves}}]{DellaTorre2015}
{Della Torre}, S., {Gervasi}, M., {Rancoita}, P.~G., {Rozza}, D., \& {Treves},
  A. 2015, Journal of High Energy Astrophysics, 8, 27,
  \dodoi{10.1016/j.jheap.2015.08.001}

\bibitem[{{Dodson} {et~al.}(2003){Dodson}, {Lewis}, {McConnell}, \&
  {Deshpande}}]{Dodson2003}
{Dodson}, R., {Lewis}, D., {McConnell}, D., \& {Deshpande}, A.~A. 2003, \mnras,
  343, 116, \dodoi{10.1046/j.1365-8711.2003.06653.x}

\bibitem[{{Dubner} {et~al.}(2017){Dubner}, {Castelletti}, {Kargaltsev},
  {Pavlov}, {Bietenholz}, \& {Talavera}}]{Dubner2017}
{Dubner}, G., {Castelletti}, G., {Kargaltsev}, O., {et~al.} 2017, \apj, 840,
  82, \dodoi{10.3847/1538-4357/aa6983}

\bibitem[{{Freeman} {et~al.}(2001){Freeman}, {Doe}, \&
  {Siemiginowska}}]{Freeman2001}
{Freeman}, P., {Doe}, S., \& {Siemiginowska}, A. 2001, in Society of
  Photo-Optical Instrumentation Engineers (SPIE) Conference Series, Vol. 4477,
  Astronomical Data Analysis, ed. J.-L. {Starck} \& F.~D. {Murtagh}, 76--87,
  \dodoi{10.1117/12.447161}

\bibitem[{{Fruscione} {et~al.}(2006){Fruscione}, {McDowell}, {Allen},
  {Brickhouse}, {Burke}, {Davis}, {Durham}, {Elvis}, {Galle}, {Harris},
  {Huenemoerder}, {Houck}, {Ishibashi}, {Karovska}, {Nicastro}, {Noble},
  {Nowak}, {Primini}, {Siemiginowska}, {Smith}, \& {Wise}}]{Fruscione2006}
{Fruscione}, A., {McDowell}, J.~C., {Allen}, G.~E., {et~al.} 2006, in Society
  of Photo-Optical Instrumentation Engineers (SPIE) Conference Series, Vol.
  6270, Society of Photo-Optical Instrumentation Engineers (SPIE) Conference
  Series, ed. D.~R. {Silva} \& R.~E. {Doxsey}, 62701V,
  \dodoi{10.1117/12.671760}

\bibitem[{Gaensler \& Slane(2006)}]{Gaensler2006}
Gaensler, B.~M., \& Slane, P.~O. 2006, Annual Review of Astronomy and
  Astrophysics, 44, 17, \dodoi{10.1146/annurev.astro.44.051905.092528}

\bibitem[{Halpern {et~al.}(2008)Halpern, Camilo, Giuliani, Gotthelf,
  McLaughlin, Mukherjee, Pellizzoni, Ransom, Roberts, \& Tavani}]{Halpern2008}
Halpern, J.~P., Camilo, F., Giuliani, A., {et~al.} 2008, The Astrophysical
  Journal, \dodoi{10.1086/594117}

\bibitem[{{Helfand} {et~al.}(2001){Helfand}, {Gotthelf}, \&
  {Halpern}}]{Helfand2001}
{Helfand}, D.~J., {Gotthelf}, E.~V., \& {Halpern}, J.~P. 2001, \apj, 556, 380,
  \dodoi{10.1086/321533}

\bibitem[{Hessels {et~al.}(2004)Hessels, Roberts, Ransom, Kaspi, Romani, Ng,
  Freire, \& Gaensler}]{Hessels2004}
Hessels, J. W.~T., Roberts, M. S.~E., Ransom, S.~M., {et~al.} 2004, The
  Astrophysical Journal, \dodoi{10.1086/422408}

\bibitem[{{Kargaltsev} \& {Pavlov}(2004)}]{2004IAUS..218..195K}
{Kargaltsev}, O., \& {Pavlov}, G. 2004, {in IAU Symposium, Vol. 218, Young
  Neutron Stars and Their Environments}, ed. F.~{Camilo} \& B.~M. {Gaensler},
  195.
\newblock \doarXiv{astro-ph/0310767}

\bibitem[{{Kargaltsev} \& {Pavlov}(2008)}]{Kargaltsev2008}
{Kargaltsev}, O., \& {Pavlov}, G.~G. 2008, in American Institute of Physics
  Conference Series, Vol. 983, 40 Years of Pulsars: Millisecond Pulsars,
  Magnetars and More, ed. C.~{Bassa}, Z.~{Wang}, A.~{Cumming}, \& V.~M.
  {Kaspi}, 171--185, \dodoi{10.1063/1.2900138}

\bibitem[{{Kargaltsev} {et~al.}(2012){Kargaltsev}, {Pavlov}, \&
  {Durant}}]{Kargaltsev2012}
{Kargaltsev}, O., {Pavlov}, G.~G., \& {Durant}, M. 2012, in Astronomical
  Society of the Pacific Conference Series, Vol. 466, Electromagnetic Radiation
  from Pulsars and Magnetars, ed. W.~{Lewandowski}, O.~{Maron}, \& J.~{Kijak},
  167.
\newblock \doarXiv{1207.1681}

\bibitem[{{Kargaltsev} {et~al.}(2017){Kargaltsev}, {Pavlov}, {Klingler}, \&
  {Rangelov}}]{Kargaltsev2017}
{Kargaltsev}, O., {Pavlov}, G.~G., {Klingler}, N., \& {Rangelov}, B. 2017,
  Journal of Plasma Physics, 83, 635830501, \dodoi{10.1017/S0022377817000630}

\bibitem[{{Kirichenko} {et~al.}(2015){Kirichenko}, {Danilenko}, {Shternin},
  {Shibanov}, {Ryspaeva}, {Zyuzin}, {Durant}, {Kargaltsev}, {Pavlov}, \&
  {Cabrera-Lavers}}]{Kirichenko2015}
{Kirichenko}, A., {Danilenko}, A., {Shternin}, P., {et~al.} 2015, \apj, 802,
  17, \dodoi{10.1088/0004-637X/802/1/17}

\bibitem[{{Klingler} {et~al.}(2018){Klingler}, {Kargaltsev}, {Pavlov}, {Ng},
  {Beniamini}, \& {Volkov}}]{Klingler2018}
{Klingler}, N., {Kargaltsev}, O., {Pavlov}, G.~G., {et~al.} 2018, \apj, 861, 5,
  \dodoi{10.3847/1538-4357/aac6e0}

\bibitem[{{Krassilchtchikov} {et~al.}(2017){Krassilchtchikov}, {Bykov},
  {Castelletti}, {Dubner}, {Kargaltsev}, \& {Pavlov}}]{Krassilchtchikov2017}
{Krassilchtchikov}, A.~M., {Bykov}, A.~M., {Castelletti}, G.~M., {et~al.} 2017,
  in Journal of Physics Conference Series, Vol. 798, Journal of Physics
  Conference Series, 012003, \dodoi{10.1088/1742-6596/798/1/012003}

\bibitem[{{Liu} {et~al.}(2022){Liu}, {Ng}, \& {Dodson}}]{Liu2022}
{Liu}, Y.~H., {Ng}, C.-Y., \& {Dodson}, R. 2022, \apj

\bibitem[{{McMullin} {et~al.}(2007){McMullin}, {Waters}, {Schiebel}, {Young},
  \& {Golap}}]{2007ASPC..376..127M}
{McMullin}, J.~P., {Waters}, B., {Schiebel}, D., {Young}, W., \& {Golap}, K.
  2007, in Astronomical Society of the Pacific Conference Series, Vol. 376,
  Astronomical Data Analysis Software and Systems XVI, ed. R.~A. {Shaw},
  F.~{Hill}, \& D.~J. {Bell}, 127

\bibitem[{Mizuno {et~al.}(2017)Mizuno, Tanaka, Takahashi, Katsuta, Hayashi, \&
  Yamazaki}]{Mizuno2017}
Mizuno, T., Tanaka, N., Takahashi, H., {et~al.} 2017, The Astrophysical
  Journal, \dodoi{10.3847/1538-4357/aa7201}

\bibitem[{{Ng} {et~al.}(2017){Ng}, {Bandiera}, {Hunstead}, \&
  {Johnston}}]{Ng2017}
{Ng}, C.~Y., {Bandiera}, R., {Hunstead}, R.~W., \& {Johnston}, S. 2017, \apj,
  842, 100, \dodoi{10.3847/1538-4357/aa762e}

\bibitem[{{Ng} {et~al.}(2012){Ng}, {Bucciantini}, {Gaensler}, {Camilo},
  {Chatterjee}, \& {Bouchard}}]{ng2012}
{Ng}, C.~Y., {Bucciantini}, N., {Gaensler}, B.~M., {et~al.} 2012, \apj, 746,
  105, \dodoi{10.1088/0004-637X/746/1/105}

\bibitem[{{Ng} \& {Romani}(2004)}]{Ng2004}
{Ng}, C.~Y., \& {Romani}, R.~W. 2004, \apj, 601, 479, \dodoi{10.1086/380486}

\bibitem[{{Ng} \& {Romani}(2008)}]{NgRomani2008}
---. 2008, \apj, 673, 411, \dodoi{10.1086/523935}

\bibitem[{{Olmi} \& {Bucciantini}(2019)}]{Olmi2019}
{Olmi}, B., \& {Bucciantini}, N. 2019, \mnras, 488, 5690,
  \dodoi{10.1093/mnras/stz2089}

\bibitem[{{Olmi} {et~al.}(2016){Olmi}, {Del Zanna}, {Amato}, {Bucciantini}, \&
  {Mignone}}]{Olmi2016}
{Olmi}, B., {Del Zanna}, L., {Amato}, E., {Bucciantini}, N., \& {Mignone}, A.
  2016, Journal of Plasma Physics, 82, 635820601,
  \dodoi{10.1017/S0022377816000957}

\bibitem[{{Pacholczyk}(1970)}]{Pacholczyk1970}
{Pacholczyk}, A.~G. 1970, {Radio astrophysics. Nonthermal processes in galactic
  and extragalactic sources}

\bibitem[{Perley \& Butler(2013)}]{Perley2013}
Perley, R.~A., \& Butler, B.~J. 2013, The Astrophysical Journal Supplement
  Series, 206, 16, \dodoi{10.1088/0067-0049/206/2/16}

\bibitem[{Perley \& Butler(2017)}]{Perley2017}
---. 2017, The Astrophysical Journal Supplement Series, 230, 7,
  \dodoi{10.3847/1538-4365/aa6df9}

\bibitem[{{Ponomaryov} {et~al.}(2018){Ponomaryov}, {Levenfish},
  {Krassilchtchikov}, {Kropotina}, \& {Petrov}}]{Ponomaryov2018}
{Ponomaryov}, G.~A., {Levenfish}, K.~P., {Krassilchtchikov}, A.~M.,
  {Kropotina}, Y.~A., \& {Petrov}, A.~E. 2018, in Journal of Physics Conference
  Series, Vol. 1038, Journal of Physics Conference Series, 012013,
  \dodoi{10.1088/1742-6596/1038/1/012013}

\bibitem[{{Porth} {et~al.}(2017){Porth}, {Buehler}, {Olmi}, {Komissarov},
  {Lamberts}, {Amato}, {Yuan}, \& {Rudy}}]{Porth2017}
{Porth}, O., {Buehler}, R., {Olmi}, B., {et~al.} 2017, \ssr, 207, 137,
  \dodoi{10.1007/s11214-017-0344-x}

\bibitem[{{Reynolds} {et~al.}(2017){Reynolds}, {Pavlov}, {Kargaltsev},
  {Klingler}, {Renaud}, \& {Mereghetti}}]{Reynolds2017}
{Reynolds}, S.~P., {Pavlov}, G.~G., {Kargaltsev}, O., {et~al.} 2017, \ssr, 207,
  175, \dodoi{10.1007/s11214-017-0356-6}

\bibitem[{Roberts {et~al.}(2002)Roberts, Hessels, Ransom, Kaspi, Freire,
  Crawford, \& Lorimer}]{Roberts2002}
Roberts, M. S.~E., Hessels, J. W.~T., Ransom, S.~M., {et~al.} 2002, The
  Astrophysical Journal, \dodoi{10.1086/344082}

\bibitem[{{Romani} {et~al.}(2005){Romani}, {Ng}, {Dodson}, \&
  {Brisken}}]{Romani2005}
{Romani}, R.~W., {Ng}, C.~Y., {Dodson}, R., \& {Brisken}, W. 2005, \apj, 631,
  480, \dodoi{10.1086/432527}

\bibitem[{{Sault} {et~al.}(1995){Sault}, {Teuben}, \& {Wright}}]{Sault1995}
{Sault}, R.~J., {Teuben}, P.~J., \& {Wright}, M.~C.~H. 1995, in Astronomical
  Society of the Pacific Conference Series, Vol.~77, Astronomical Data Analysis
  Software and Systems IV, ed. R.~A. {Shaw}, H.~E. {Payne}, \& J.~J.~E.
  {Hayes}, 433.
\newblock \doarXiv{astro-ph/0612759}

\bibitem[{{Saz Parkinson} {et~al.}(2010){Saz Parkinson}, {Dormody}, {Ziegler},
  {Ray}, {Abdo}, {Ballet}, {Baring}, {Belfiore}, {Burnett}, {Caliandro},
  {Camilo}, {Caraveo}, {de Luca}, {Ferrara}, {Freire}, {Grove}, {Gwon},
  {Harding}, {Johnson}, {Johnson}, {Johnston}, {Keith}, {Kerr},
  {Kn{\"o}dlseder}, {Makeev}, {Marelli}, {Michelson}, {Parent}, {Ransom},
  {Reimer}, {Romani}, {Smith}, {Thompson}, {Watters}, {Weltevrede}, {Wolff}, \&
  {Wood}}]{SazParkinson2010}
{Saz Parkinson}, P.~M., {Dormody}, M., {Ziegler}, M., {et~al.} 2010, \apj, 725,
  571, \dodoi{10.1088/0004-637X/725/1/571}

\bibitem[{{Strom} \& {Greidanus}(1992)}]{Strom&Greidanus1992}
{Strom}, R.~G., \& {Greidanus}, H. 1992, \nat, 358, 654,
  \dodoi{10.1038/358654a0}

\bibitem[{{Toropina} {et~al.}(2019){Toropina}, {Romanova}, \&
  {Lovelace}}]{Toropina2019}
{Toropina}, O.~D., {Romanova}, M.~M., \& {Lovelace}, R.~V.~E. 2019, \mnras,
  484, 1475, \dodoi{10.1093/mnras/stz034}

\bibitem[{Van~Etten {et~al.}(2008)Van~Etten, Romani, \& Ng}]{VanEtten2008}
Van~Etten, A., Romani, R.~W., \& Ng, C. 2008, The Astrophysical Journal, 680,
  1417, \dodoi{10.1086/587865}

\bibitem[{{Van Etten} {et~al.}(2012){Van Etten}, {Romani}, \&
  {Ng}}]{VanEtten2012}
{Van Etten}, A., {Romani}, R.~W., \& {Ng}, C.~Y. 2012, \apj, 755, 151,
  \dodoi{10.1088/0004-637X/755/2/151}

\bibitem[{{Yao} {et~al.}(2017){Yao}, {Manchester}, \& {Wang}}]{Yao2017}
{Yao}, J.~M., {Manchester}, R.~N., \& {Wang}, N. 2017, \apj, 835, 29,
  \dodoi{10.3847/1538-4357/835/1/29}

\bibitem[{{Yusef-Zadeh} \& {Gaensler}(2005)}]{Yusef2005}
{Yusef-Zadeh}, F., \& {Gaensler}, B.~M. 2005, Advances in Space Research, 35,
  1129, \dodoi{10.1016/j.asr.2005.03.003}

\end{thebibliography}
\bibliographystyle{aasjournal}



\end{document}